\documentclass[aps,prd,twocolumn,nofootinbib,10pt]{revtex4-2}
\usepackage{amsmath,amssymb}
\usepackage{graphicx}
\usepackage{hyperref}
\usepackage{orcidlink}

\begin{document}

\title{Shadow, Quasinormal Modes, Sparsity, and Energy Emission Rate of Euler-Heisenberg Black Hole Surrounded by Perfect Fluid Dark Matter}

\author{Edilberto~O.~Silva\orcidlink{0000-0002-0297-5747}}
\email{edilberto.silva@ufma.br}
\affiliation{Programa de P\'{o}s-Gradua\c{c}\~{a}o em F\'{i}sica \& Coordena\c c\~ao do Curso de F\'{\i}sica -- Bacharelado, Universidade Federal do Maranh\~{a}o, 65080-805, S\~{a}o Lu\'{i}s, Maranh\~{a}o, Brazil}

\author{Faizuddin Ahmed\,\orcidlink{0000-0003-2196-9622}}
\email{faizuddinahmed15@gmail.com (Corresponding author)}
\affiliation{Department of Physics, The Assam Royal Global University, Guwahati, Assam 781035, India}

\begin{abstract}
In this work, we investigate the optical, dynamical, and radiative properties of an Euler--Heisenberg black hole immersed in a perfect fluid dark matter (PFDM) background. We analyze the photon sphere and shadow, the scalar quasinormal-mode spectrum in the eikonal regime, the grey-body factor through the eikonal QNM correspondence, the sparsity of Hawking radiation, and the corresponding energy emission rate. Our results show that both the black-hole charge and the PFDM parameter significantly affect the photon sphere, shadow size, quasinormal frequencies, Hawking temperature, and emission profile, whereas the Euler--Heisenberg correction is typically subleading in the parameter range explored, although it may become more visible in strong-charge regimes for selected observables. Overall, the dark-matter environment provides the dominant imprint on the phenomenology of the system, indicating that shadow and ringdown-related quantities may serve as useful probes of PFDM effects within the approximations considered.
\end{abstract}


\maketitle

%\tableofcontents

\section{Introduction}\label{sec:1}

Black holes provide a natural laboratory for probing gravity, quantum effects, and environmental interactions in the strong-field regime. Their thermodynamic interpretation, established through the seminal works of Bekenstein and Hawking, revealed that horizon area and surface gravity admit an entropy--temperature description, thereby opening a broad line of research on black-hole radiation, stability, and quantum aspects of gravitation \cite{Bekenstein1973,Hawking1975,HawkingPage1983}. In parallel, the response of black holes to external perturbations led to the theory of quasinormal modes (QNMs), which now plays a central role in black-hole spectroscopy and in the interpretation of gravitational-wave ringdown signals \cite{ReggeWheeler1957,Zerilli1970,Vishveshwara1970,KokkotasSchmidt1999,BertiCardosoStarinets2009}.

Extensions of the standard charged black-hole scenario are particularly relevant when one aims to incorporate quantum corrections to the electromagnetic sector. Among the most important examples, Euler--Heisenberg electrodynamics arises as an effective description of quantum-electrodynamic vacuum polarization in strong electromagnetic fields \cite{HeisenbergEuler1936,Plebanski1970,SalazarGarciaPlebanski1987}. When coupled to gravity, it modifies the spacetime geometry and the propagation of light, thereby affecting both optical and dynamical observables associated with black holes \cite{YajimaTamaki2001,BretonLopez2021}. Such corrections offer a useful framework for exploring deviations from the Reissner--Nordstr\"om picture while remaining close to physically motivated effective field theory.

Besides intrinsic modifications of the black-hole geometry, environmental effects are also expected to leave observable signatures. In this context, perfect fluid dark matter (PFDM) provides a simple and widely used phenomenological model that introduces a logarithmic correction to the metric and has been employed to investigate geodesic structure, shadows, accretion, oscillation spectra, and other strong-gravity observables \cite{RahamanEtAl2010,LiYang2012,HaroonEtAl2019,Ahmed2025, Shahzad2025,Ma2024, Ahmed2026ChargedBardeen,Ahmed2026DarkMatterStringCloud, AlBadawi2026BlackHoleStudy,Ahmed2026ChargedBHNonlinear}. Since realistic astrophysical black holes are not isolated systems, understanding how surrounding matter distributions alter observable quantities is essential for connecting theoretical models with data.

The study of black-hole shadows has gained special relevance in recent years. The foundations of the subject go back to the early works of Synge, Cunningham and Bardeen, and Luminet, while modern treatments have refined the use of shadow observables as probes of the near-horizon geometry and of matter fields around compact objects \cite{Synge1966,CunninghamBardeen1973,Luminet1979,HiokiMaeda2009,PerlickTsupko2022}. This field received decisive observational impetus from the Event Horizon Telescope (EHT), which produced horizon-scale images of M87* and Sgr A*, turning black-hole shadow phenomenology into a directly testable branch of relativistic astrophysics \cite{EHTM87I2019,EHTM87VI2019,EHTSgrAI2022,EHTSgrAVI2022}. In parallel, QNMs provide complementary information about the dynamical response of black holes, and in the eikonal regime they are closely connected with the properties of unstable circular null geodesics \cite{SchutzWill1985,IyerWill1987,CardosoEtAl2009}. This geodesic correspondence makes shadow observables and QNM spectra especially suitable for a unified analysis.

Hawking radiation offers a further window into the quantum behavior of black holes. Beyond the standard characterization in terms of temperature and spectral emission rates \cite{Page1976a,Page1976b}, recent investigations have emphasized that the Hawking cascade is typically sparse rather than continuous, and that this sparsity depends sensitively on the underlying geometry and matter content \cite{GrayEtAl2015,ChowdhuryBanerjee2020}. Likewise, the relation between shadow size, absorption cross-section, grey-body factors, and energy emission rate has become increasingly useful for linking optical and radiative properties within a common framework \cite{WeiLiu2011,WeiLiu2013,DecaniniEtAl2011,KonoplyaZhidenko2024,KonoplyaZhidenko2025}.

Motivated by these developments, in this work we study the shadow, scalar perturbations in the eikonal limit, the associated QNM spectrum, the eikonal grey-body factor, the sparsity of Hawking radiation, and the energy emission rate of an Euler--Heisenberg black hole surrounded by PFDM. Related aspects of Euler--Heisenberg black holes and dark-matter backgrounds have been considered in several previous contexts. For example, the shadow and QNMs of rotating Einstein--Euler--Heisenberg black holes were investigated in \cite{LambiaseEtAl2025}, while QNM properties of charged black holes in PFDM backgrounds were analyzed in \cite{TanEtAl2025}, and quasi-periodic oscillations in charged black holes with PFDM were explored in \cite{Ahmed2026ChargedRegularBH}. However, a combined analysis of optical, dynamical, and radiative observables for the Euler--Heisenberg--PFDM system remains comparatively unexplored.

Our goal is therefore to provide a unified picture of how the Euler--Heisenberg parameter and the PFDM background jointly influence the photon sphere, shadow radius, quasinormal frequencies, grey-body factors, Hawking-radiation sparsity, and energy emission spectrum. In particular, we aim to identify which observables are predominantly controlled by the dark-matter environment and which may retain sensitivity to nonlinear electromagnetic corrections. As will be shown, the PFDM parameter generally produces the dominant phenomenological imprint in the parameter domain considered, whereas the Euler--Heisenberg correction typically enters as a secondary effect, although it can become more visible in selected high-charge configurations.

\section{Geometric Background: BH Metric with PFDM }\label{sec:2}

The total action for black holes with EH--PFDM spacetime can be given~\cite{Ma2024} 
\begin{equation} 
S = \int d^4 x \sqrt{-g} \left[ \frac{R}{16\pi} + \mathcal{L}_{\mathrm{EH}}(F) + \mathcal{L}_{\mathrm{PFDM}} \right],\label{aa1} 
\end{equation} 
where $R$ is the Ricci scalar, $\mathcal{L}_{\mathrm{EH}}$ denotes the EH nonlinear electromagnetic Lagrangian, and $\mathcal{L}_{\mathrm{PFDM}}$ represents the effective Lagrangian of perfect fluid dark matter. The electromagnetic invariant is defined as $F \equiv \frac{1}{4} F_{\mu\nu}\,F^{\mu\nu}$. The EH Lagrangian up to leading-order QED correction takes the form~\cite{Heisenberg1936,Yajima2001,Plebanski1970,Salazar1987} 
\begin{equation} 
\mathcal{L}_{\mathrm{EH}}(F) = -\frac{1}{4}\,F + \frac{\alpha}{8}\,F^2,\label{aa2}
\end{equation} 
where the parameter $\alpha$ encodes the strength of quantum electrodynamic corrections. 

Varying the action (\ref{aa1}) with respect to the metric yields the Einstein field equations, 
\begin{equation} 
G_{\mu\nu} = 8\pi \left( T_{\mu\nu}^{\mathrm{EH}} + T_{\mu\nu}^{\mathrm{PFDM}} \right),\label{aa3}
\end{equation} 
where the total energy-momentum tensor is the sum of contributions from nonlinear electrodynamics and dark matter.

We assume a static and spherically symmetric spacetime characterized by the line element
\begin{equation}
ds^{2} = - f(r)\, dt^{2} + \frac{1}{f(r)}\, dr^{2} + r^{2} d\Omega^{2},\label{metric}
\end{equation}
where $d\Omega^{2} = d\theta^{2} + \sin^{2}\theta\, d\phi^{2}$ defines the metric of the unit two-sphere. 

For the EH electromagnetic field, the energy-momentum tensor in the purely electric case is given by Magos and Breton~\cite{Magos2020}
\begin{equation}
T^{t}_{\ t\,(\mathrm{EH})} = \frac{1}{4\pi} \left( \frac{Q^2}{2r^4} + \frac{\alpha Q^4}{8r^8} \right),\label{aa4}
\end{equation}
where $Q$ is the electric charge of the black hole.

For perfect fluid dark matter, the effective energy-momentum tensor reads~\cite{Li2012}
\begin{equation}
T^{t}_{\ t\,(\mathrm{PFDM})} = \frac{\lambda}{8\pi r^3},\label{aa5}
\end{equation}
where $\lambda$ is the dark matter parameter characterizing the density profile.

Substituting Eqs. (\ref{aa4}) and (\ref{aa5}) into the Einstein equations (\ref{aa3}), the metric function $f(r)$ takes the explicit form:
\begin{equation}
f(r) = 1 -\frac{2M}{r} + \frac{Q^{2}}{r^{2}} - \frac{\alpha Q^{4}}{20\, r^{6}}+\frac{\lambda}{r} \ln\!\frac{r}{|\lambda|}.\label{function}
\end{equation}
For a large radial coordinate $r$, the asymptotic expansion is given by
\begin{equation}
\lim_{r \to \infty} f(r) =1,
\end{equation}
which shows asymptotic flatness of the spacetime. 

In the limit $\alpha \to 0$, corresponding to the absence of EH effects, the considered space-time simplifies to Reissner-Nordstrom black hole surrounded by PFDM \cite{Xu2017}. Moreover, in the limit $Q \to 0$, the space-time simplifies to the Schwrazschild black hole with PFDM, reported in \cite{Li2012}

\section{Black Hole Shadow}\label{sec:3}

The black hole shadow is the dark region produced by strong gravitational lensing of light around the event horizon, where photons are captured or deflected into unstable orbits. Observations from EHT, particularly for M87* and Sagittarius A*, have provided the first direct images of these shadows, offering a powerful test of general relativity in the strong-field regime.

The null geodesic condition is given by
\begin{equation}
    ds^2=0 \Longrightarrow g_{\mu\nu} \dot x^{\mu} \dot x^{\nu}=0,\label{bb1}
\end{equation}
where the dot denotes the derivative with respect to an affine parameter $\tau$.

Using metric (\ref{metric}), we find
\begin{equation}
    -f(r) \dot t^2+\frac{\dot r^2}{f(r)}+r^2 \dot \theta^2+r^2 \sin^2 \theta \dot \phi^2=0.\label{bb2}
\end{equation}

Using the relations \(\dot t=E/f(r)\) and \(\dot \phi=L/r^2\), where $E$ and $L$ are the conserved energy and the angular momentum, respectively, the equation of motion for photon particles is obtained as
\begin{equation}
    \dot r^2+V_{\rm eff}(r)=E^2,\label{bb4}
\end{equation}
where $V_{\rm eff}=\frac{L^2}{r^2}\,f(r)$ is the effective potential governing the dynamics of massless particles.

For circular null orbits, the conditions $\dot r=0$ and $\ddot r=0$ must be satisfied. Using (\ref{bb4}), we find
\begin{equation}
    \frac{\partial V_{\rm eff}}{\partial r}=0.\label{bb7}
\end{equation}
This relation gives the photon sphere radius satisfying the following polynomial relation in $r$:
\begin{equation}
1 - \frac{3M}{r_p} + \frac{2Q^2}{r^2_p}- \frac{\alpha Q^4}{5r^6_p}+ \frac{3\lambda}{2r_p}\ln\!\frac{r_p}{|\lambda|}- \frac{\lambda}{2r_p} = 0.\label{bb8}
\end{equation}
The exact analytical solution of the above equation yields the photon sphere radius $r_p$. However, due to the presence of the logarithmic term, an exact closed-form solution is not feasible. Nevertheless, by specifying suitable values of the geometric parameters, the photon sphere radius can be determined numerically with high precision.

As the considered spacetime is asymptotically flat, the shadow radius equals the critical impact parameter for photon particles at radius $r=r_p$. We find the shadow expression as follows \cite{PerlickTsupko2022}:
\begin{align}
R_{\rm sh}=\beta_c&=\frac{r_p}{\sqrt{f(r_p)}}\nonumber\\
&=\frac{r_p}{\sqrt{1 - \frac{2M}{r_p} + \frac{Q^{2}}{r^{2}_p} - \frac{\alpha Q^{4}}{20\, r^{6}_p} + \frac{\lambda}{r_p}\ln\!\frac{r_p}{|\lambda|}}}.\label{bb9}
\end{align}

\begin{table}[ht!]
\centering
\begin{tabular}{|c|ccccc|}
\hline
$Q/M$ & \multicolumn{5}{c|}{$\lambda/M$} \\
\cline{2-6}
 & $-0.5$ & $-0.4$ & $-0.3$ & $-0.2$ & $-0.1$ \\
\hline
2 & 2.88800 & 2.85832 & 2.81860 & 2.76783 & 2.70291 \\
3 & 3.22266 & 3.20773 & 3.18783 & 3.16182 & 3.12720 \\
4 & 3.62115 & 3.60966 & 3.59480 & 3.57563 & 3.55020 \\
5 & 4.00128 & 3.99148 & 3.97906 & 3.96328 & 3.94249 \\
\hline
\end{tabular}
\caption{Numerical results of photon sphere radius $r_p/M$ for varying $Q/M$ and $\lambda/M$ at $\alpha/M^2=0.10 \times 10^{3}$.}
\label{tab:1}
\end{table}

\begin{table}[ht!]
\centering
\begin{tabular}{|c|ccccc|}
\hline
$Q/M$ & \multicolumn{5}{c|}{$\lambda/M$} \\
\cline{2-6}
 & -0.5 & -0.4 & -0.3 & -0.2 & -0.1 \\
\hline
2 & 3.14523 & 3.10925 & 3.06336 & 3.00641 & 2.93503 \\
3 & 3.53358 & 3.51446 & 3.48999 & 3.45896 & 3.41863 \\
4 & 3.97716 & 3.96243 & 3.94394 & 3.92070 & 3.89051 \\
5 & 4.39882 & 4.38626 & 4.37075 & 4.35146 & 4.32653 \\
\hline
\end{tabular}
\caption{Numerical results of photon sphere radius $r_p/M$ for varying $Q/M$ and $\lambda/M$ at $\alpha/M^2=0.15 \times 10^{3}$.}
\label{tab:2}
\end{table}

\begin{figure}[ht!]
\centering
\includegraphics[width=0.9\linewidth]{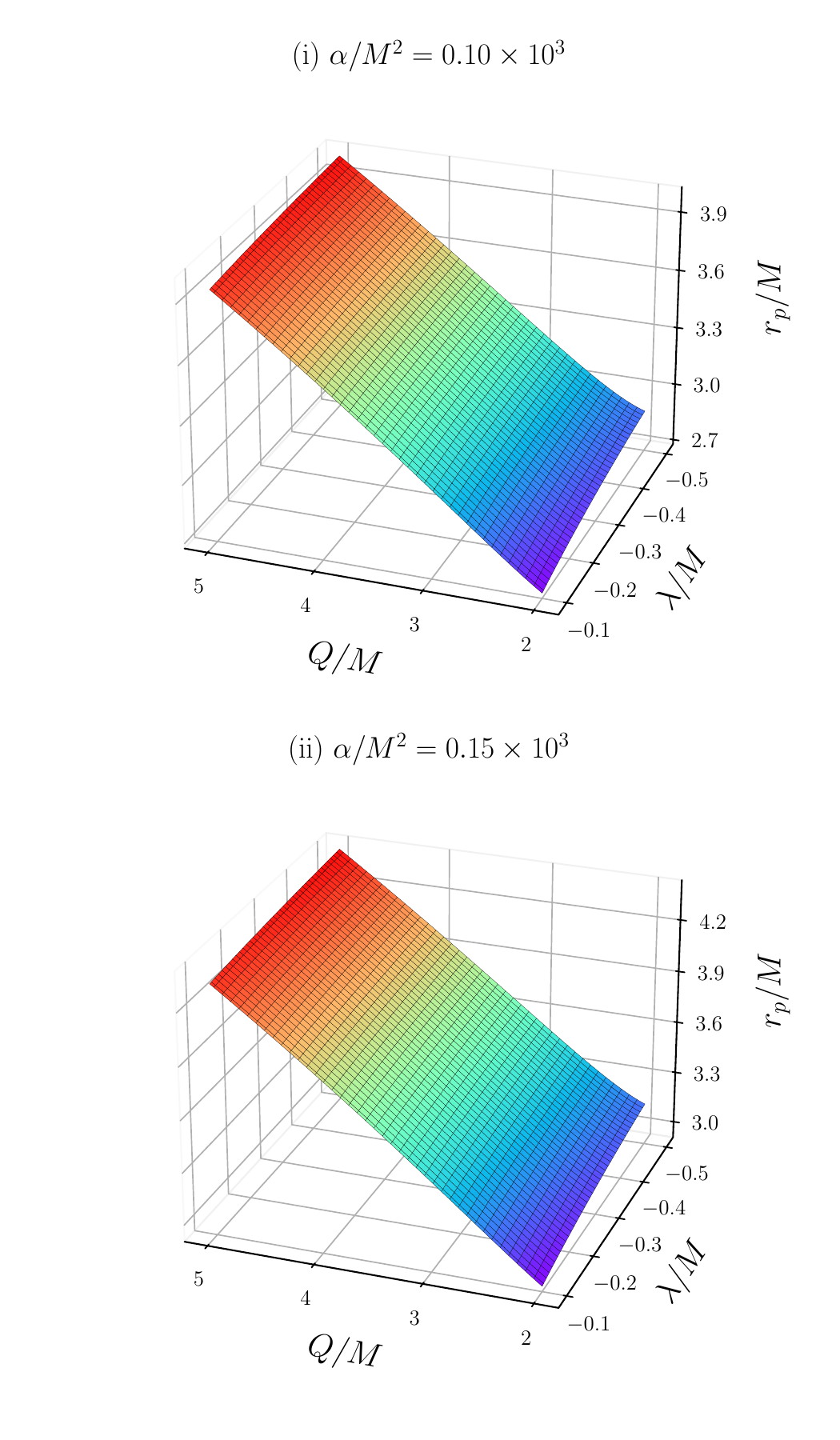}
\caption{Three-dimensional representation of the photon sphere radius as a function $(Q, \lambda)$ for two values of $\alpha$.}
\label{fig:photon-sphere}
\end{figure}

\begin{figure}[ht!]
    \centering
    \includegraphics[width=0.8\linewidth]{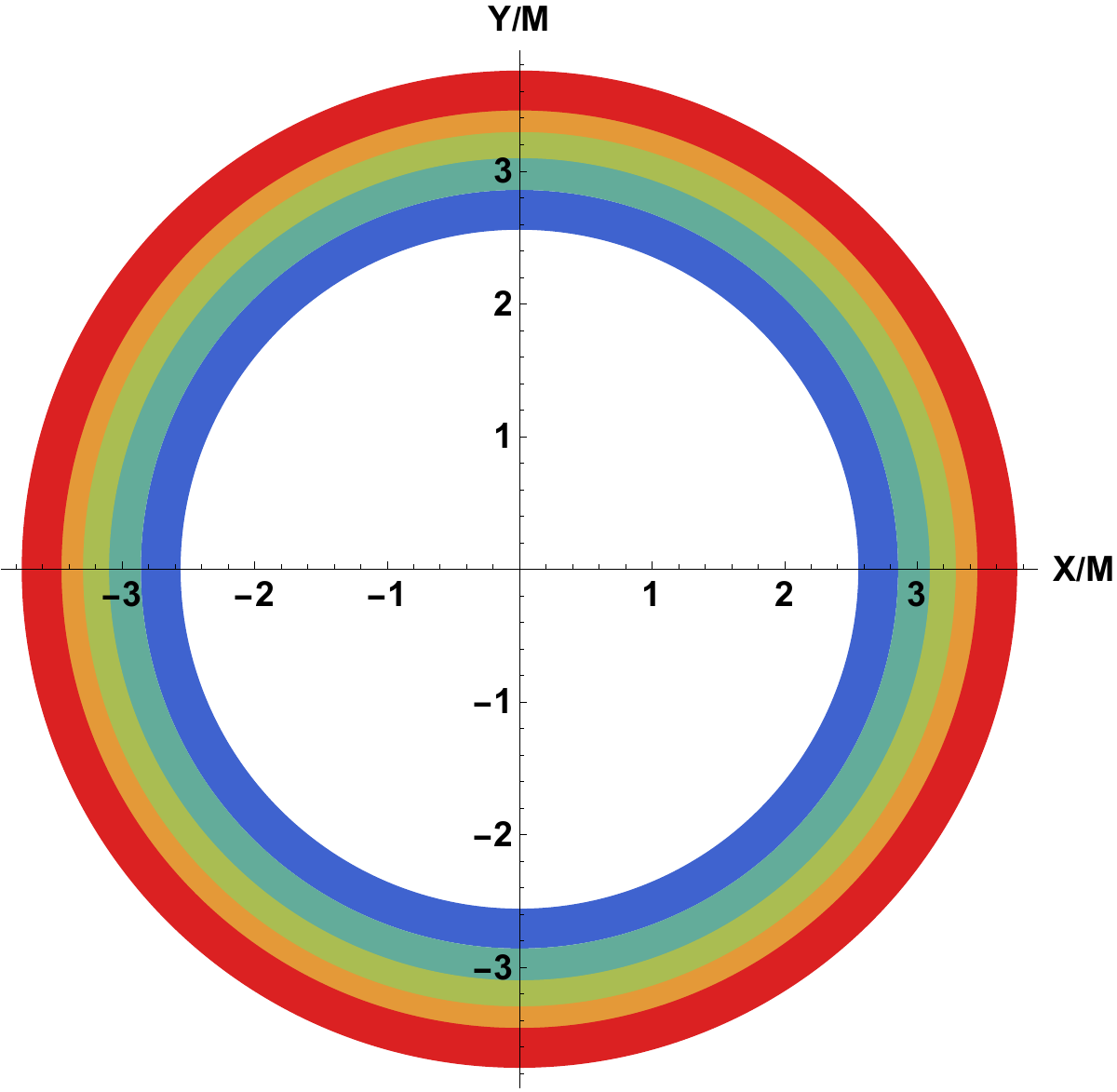}\\
    (i) $\alpha/M^2=0.10 \times 10^3$\\
    \includegraphics[width=0.8\linewidth]{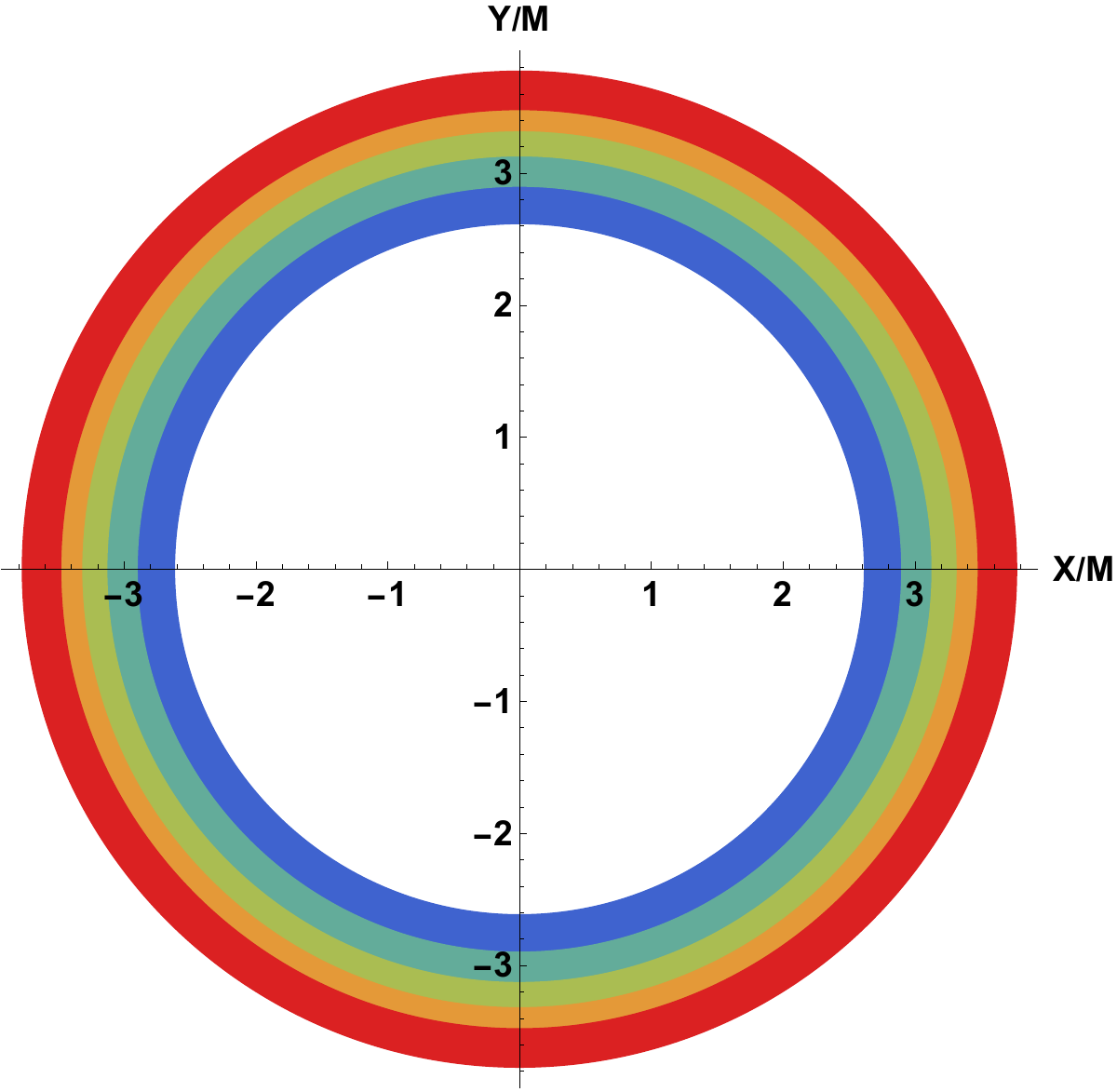}\\
    (ii) $\alpha/M^2=0.15 \times 10^3$\\
    \caption{Annular photon rings for varying $\lambda/M=-0.5$ to $\lambda/M=-0.1$ (outer to inner). Here $Q/M=1$. We observe that as EH parameter $\alpha$ increases, the size of the ring also increases which agrees with Tables \ref{tab:1}--\ref{tab:2}.}
    \label{fig:photon-rings}
\end{figure}

\begin{table}[ht!]
\centering
\begin{tabular}{|c|ccccc|}
\hline
$Q/M$ & \multicolumn{5}{c|}{$\lambda/M$} \\
\cline{2-6}
 & -0.5 & -0.4 & -0.3 & -0.2 & -0.1 \\
\hline
2 & 4.91286 & 4.71189 & 4.47942 & 4.21232 & 3.89956 \\
3 & 4.17671 & 4.07444 & 3.95423 & 3.81263 & 3.64113 \\
4 & 3.98247 & 3.91183 & 3.82906 & 3.73122 & 3.61163 \\
5 & 3.92302 & 3.86810 & 3.80408 & 3.72851 & 3.63595 \\
\hline
\end{tabular}
\caption{Numerical results of shadow radius $R_{\mathrm{sh}}/M$ for varying $Q/M$ and $\lambda/M$ at $\alpha/M^2=0.10 \times 10^{3}$.}
\label{tab:3}
\end{table}

\begin{table}[ht!]
\centering
\begin{tabular}{|c|ccccc|}
\hline
$Q/M$ & \multicolumn{5}{c|}{$\lambda/M$} \\
\cline{2-6}
 & -0.5 & -0.4 & -0.3 & -0.2 & -0.1 \\
\hline
2 & 5.30011 & 5.08493 & 4.83985 & 4.56094 & 4.23609 \\
3 & 4.69673 & 4.57605 & 4.43691 & 4.27554 & 4.08262 \\
4 & 4.55371 & 4.46723 & 4.36765 & 4.25169 & 4.11190 \\
5 & 4.52945 & 4.46084 & 4.38208 & 4.29040 & 4.17960 \\
\hline
\end{tabular}
\caption{Numerical results of shadow radius $R_{\mathrm{sh}}/M$ for varying $Q/M$ and $\lambda/M$ at $\alpha/M^2=0.15 \times 10^{3}$.}
\label{tab:4}
\end{table}

\begin{figure}[ht!]
\centering
\includegraphics[width=0.9\linewidth]{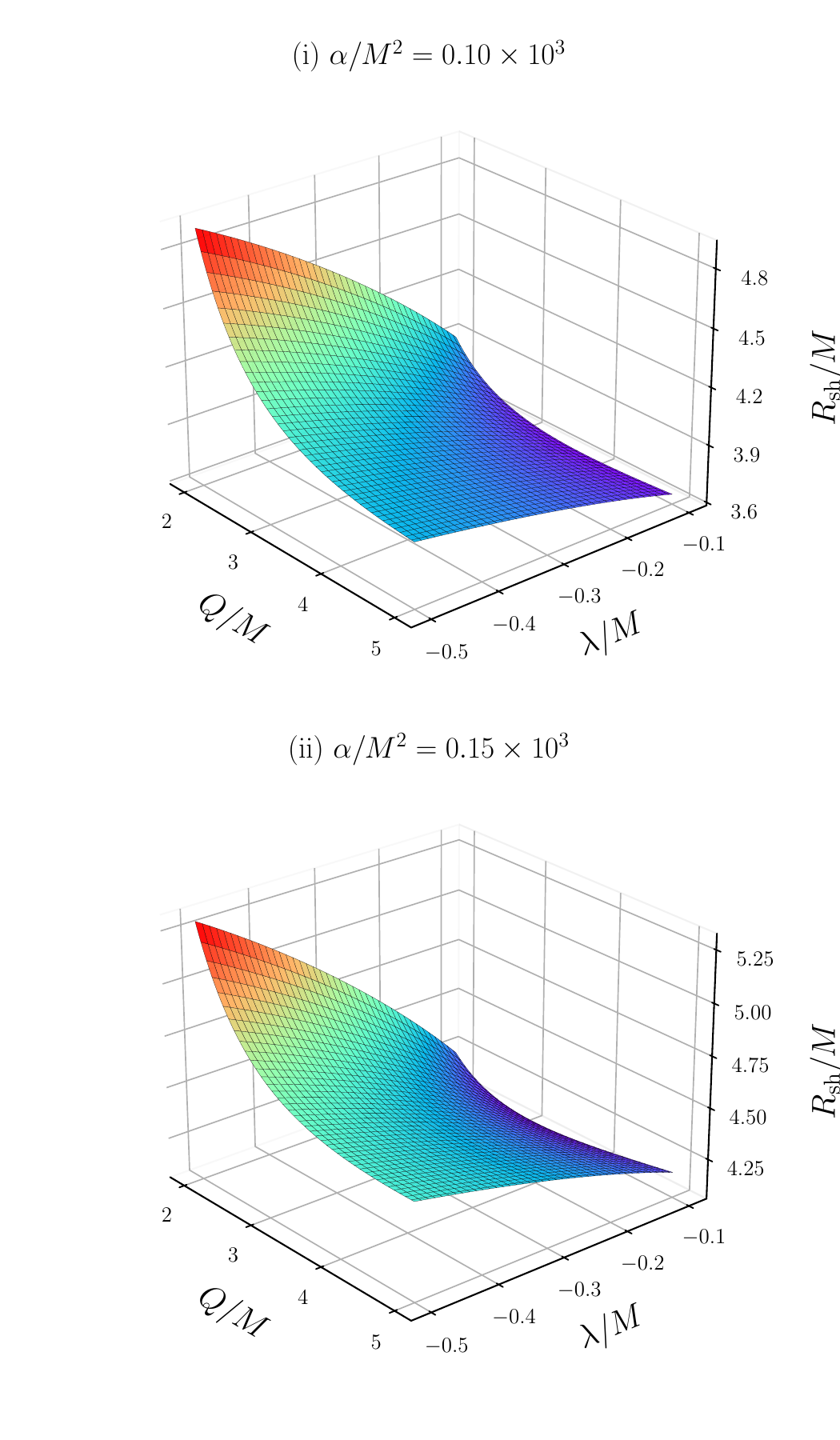}
\caption{Three-dimensional representation of the shadow radius $R_{\rm sh}$ as a function $(Q, \lambda)$ for two values of $\alpha$.}
\label{fig:shadow}
\end{figure}
\begin{figure*}[ht!]
    \centering
    \includegraphics[width=0.8\textwidth]{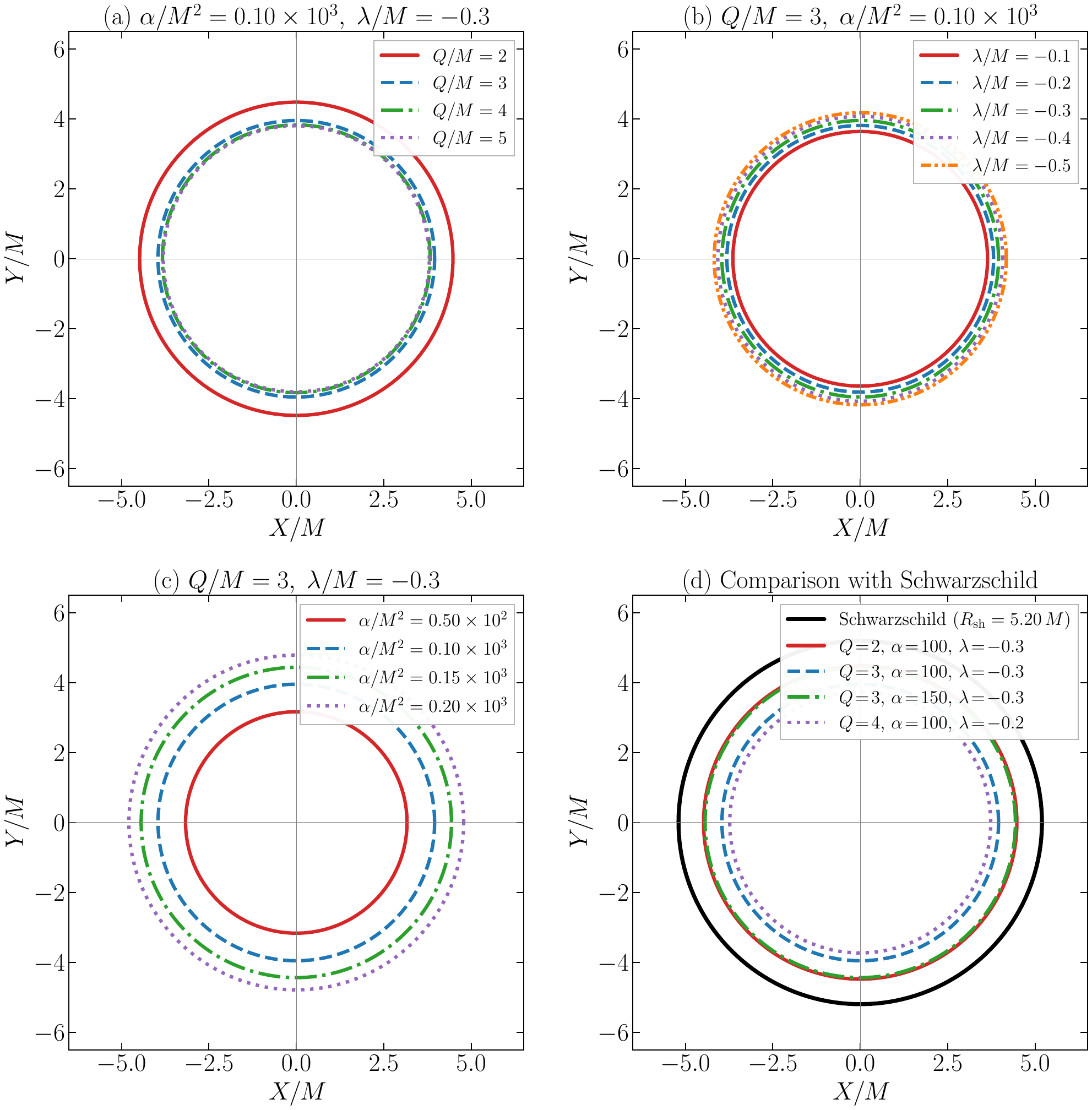}
    \caption{Shadow silhouettes of the EH--PFDM black hole in the
    observer's celestial plane.
    (a)~Varying $Q/M$ at $\alpha/M^{2}=0.10\times 10^{3}$,
    $\lambda/M=-0.3$.
    (b)~Varying $\lambda/M$ at $Q/M=3$, $\alpha/M^{2}=0.10\times 10^{3}$.
    (c)~Varying $\alpha/M^{2}$ at $Q/M=3$, $\lambda/M=-0.3$.
    (d)~Comparison of representative configurations with the
    Schwarzschild shadow (solid black circle).}
    \label{fig:shadow-sil}
\end{figure*}
The numerical values of the photon sphere radius are listed in Table~\ref{tab:1}--\ref{tab:2}, while Fig.~\ref{fig:photon-rings}--\ref{fig:photon-sphere} provides phton rings and a three-dimensional visualization of $r_p$ as a function of $(Q,\lambda)$ for two representative values of $\alpha$. Both the charge and the PFDM parameter decrease the photon sphere radius, with $\lambda$ exerting the dominant effect. The sensitivity to the EH parameter $\alpha$ is comparatively weak: the surfaces for $\alpha/M^{2}=0.1$ and $0.2$ are nearly indistinguishable, indicating that the leading-order QED correction modifies $r_p$ only at the sub-percent level in the parameter range explored.

The shadow radius is tabulated in Table~\ref{tab:3}--\ref{tab:4} and depicted in Fig.~\ref{fig:shadow}. The qualitative trends mirror those of the photon sphere: $R_{\rm sh}$ decreases monotonically with both $Q$ and $\lambda$. For small charges the reduction is modest, while at $Q/M=0.5$ and $\lambda/M=0.5$ the shadow shrinks by roughly $24\%$ relative to the weakly charged, low-$\lambda$ case. As with $r_p$, the dependence on $\alpha$ remains negligible, confirming that the shadow is primarily shaped by the dark-matter environment rather than by the nonlinear electromagnetic correction.

The shadow silhouettes projected onto the observer's celestial plane $(X,Y)$ are displayed in Fig.~\ref{fig:shadow-sil}.  Since the spacetime
is spherically symmetric, each shadow is a circle of radius $R_{\rm sh}$. Figure~\ref{fig:shadow-sil}(a) varies the electric charge at fixed $\alpha/M^{2}=0.10\times 10^{3}$ and $\lambda/M=-0.3$: increasing $Q/M$ from 2 to 5 progressively shrinks the shadow, consistent with the decrease of $r_p$ and $R_{\rm sh}$ reported in Tables~I--IV.
Figure~\ref{fig:shadow-sil}(b) varies the PFDM parameter at $Q/M=3$; the shadow contracts monotonically as $\lambda$ becomes more negative, confirming the dominant role of the dark-matter background.  In Fig.~\ref{fig:shadow-sil}(c) the EH parameter is varied at $Q/M=3$
and $\lambda/M=-0.3$: the shadow now grows substantially with $\alpha$,
demonstrating that the QED correction is clearly resolvable in the
strong-charge regime.  Finally, Fig.~\ref{fig:shadow-sil}(d) compares
selected EH--PFDM configurations with the Schwarzschild shadow
($R_{\rm sh}=3\sqrt{3}\,M\approx 5.20\,M$).  All EH--PFDM shadows lie
inside the Schwarzschild circle, quantifying the combined reduction
induced by the charge, dark matter, and nonlinear electromagnetic
correction.

\begin{figure*}[tbhp]
    \centering
    \includegraphics[width=0.8\linewidth]{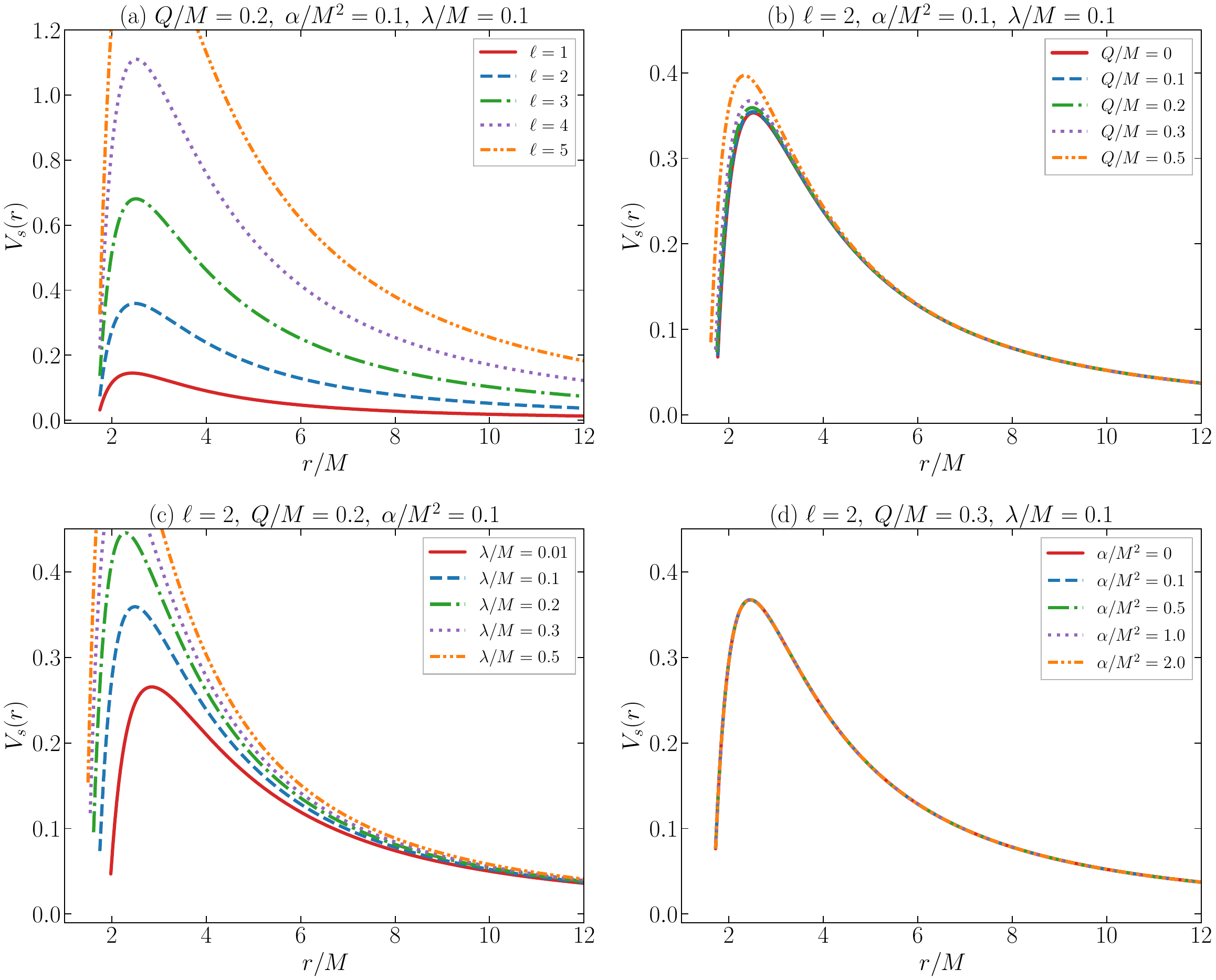}
    \caption{Effective potential $V_s(r)$ for massless scalar perturbations. (a)~Varying the multipole number $\ell$ at $Q/M=0.2$, $\alpha/M^{2}=0.1$, $\lambda/M=0.1$. (b)~Varying $Q/M$ at $\ell=2$, $\alpha/M^{2}=0.1$, $\lambda/M=0.1$. (c)~Varying $\lambda/M$ at $\ell=2$, $Q/M=0.2$, $\alpha/M^{2}=0.1$. (d)~Varying $\alpha/M^{2}$ at $\ell=2$, $Q/M=0.3$, $\lambda/M=0.1$.}
    \label{fig:potential}
\end{figure*}

\section{Scalar Perturbations}\label{sec:4}

We now consider massless scalar perturbations in the EH-PFDM background. Scalar perturbations of black holes are typically described by a minimally coupled massless scalar field satisfying the Klein--Gordon equation in a curved spacetime background. After separation of variables, the perturbation reduces to a Schr\"odinger-like wave equation with an effective potential determined by the underlying geometry. The resulting dynamics allow one to study quasinormal modes, which characterize the response of the black hole to external disturbances and provide important information about its stability and relaxation properties. Detailed discussions on scalar perturbations and quasinormal modes can be found in Refs.~\cite{KokkotasSchmidt1999,BertiCardosoStarinets2009,KonoplyaZhidenko2011,KonoplyaZhidenko2024,KonoplyaZhidenkoZinhailo2019,BertiCardosoStarinets2009}.

The scalar field $\Psi$ satisfies the Klein--Gordon equation
\begin{equation}
\frac{1}{\sqrt{-g}}\,\partial_\mu
\left(\sqrt{-g}\,g^{\mu\nu}\partial_\nu\Psi\right)=0.
\label{mm1-new}
\end{equation}
Using the standard decomposition
\begin{equation}
\Psi(t,r,\theta,\phi)=e^{-i\omega t}\,
Y_{\ell m}(\theta,\phi)\,\frac{\psi(r)}{r},
\label{mm2-new}
\end{equation}
the radial part reduces to a Schr\"odinger-like equation,
\begin{equation}
\frac{d^2\psi}{dr_*^2}+\left(\omega^2-V_s\right)\psi=0,
\label{mm3-new}
\end{equation}
where the tortoise coordinate is defined by
\begin{equation}
\frac{dr_*}{dr}=\frac{1}{f(r)}.
\label{mm4-new}
\end{equation}

The effective potential can be written in the compact form
\begin{equation}
V_s(r)=f(r)\left[\frac{\ell(\ell+1)}{r^2}+\frac{f'(r)}{r}\right]
=\frac{f(r)}{r^2}\big[\ell(\ell+1)+\Xi(r)\big],
\label{mm5-new}
\end{equation}
with
\begin{equation}
\Xi(r)=\frac{2M}{r}-\frac{2Q^2}{r^2}
+\frac{3\alpha Q^4}{10r^6}
+\frac{\lambda}{r}\left(1-\ln\!\frac{r}{|\lambda|}\right).
\label{mm5b-new}
\end{equation}
Hence, the scalar perturbation spectrum is governed by the combined effects of the EH correction and the PFDM background, as captured by the deformed effective potential.

The radial profile of $V_s(r)$ is shown in Fig.~\ref{fig:potential}. Figure~\ref{fig:potential}(a) displays the expected growth of the potential barrier with the multipole number $\ell$: higher angular momentum modes experience a taller and narrower barrier, which is the standard centrifugal effect. In Fig.~\ref{fig:potential}(b) the charge dependence is examined at fixed $\ell=2$; increasing $Q$ raises the peak height and shifts it slightly inward, consistent with the contraction of the photon sphere reported in Sec.~III. The influence of the PFDM parameter is illustrated in Fig. \ref{fig:potential}(c). Here, the effect is more pronounced: a larger $\lambda$ significantly enhances the barrier and shifts the peak to smaller radii, reflecting the deeper gravitational well induced by the dark-matter background. Finally, Fig.~\ref{fig:potential}(d) varies the EH parameter $\alpha$ at $Q/M=0.3$. The curves are barely distinguishable, confirming once again that the leading-order QED correction has a negligible impact on the scalar dynamics in the parameter regime considered. These observations establish a clear hierarchy: the PFDM environment dominates the modification of the potential barrier, while the EH correction enters only as a perturbative sub-leading effect.

\begin{figure*}[tbhp]
    \centering
    \includegraphics[width=0.8\linewidth]{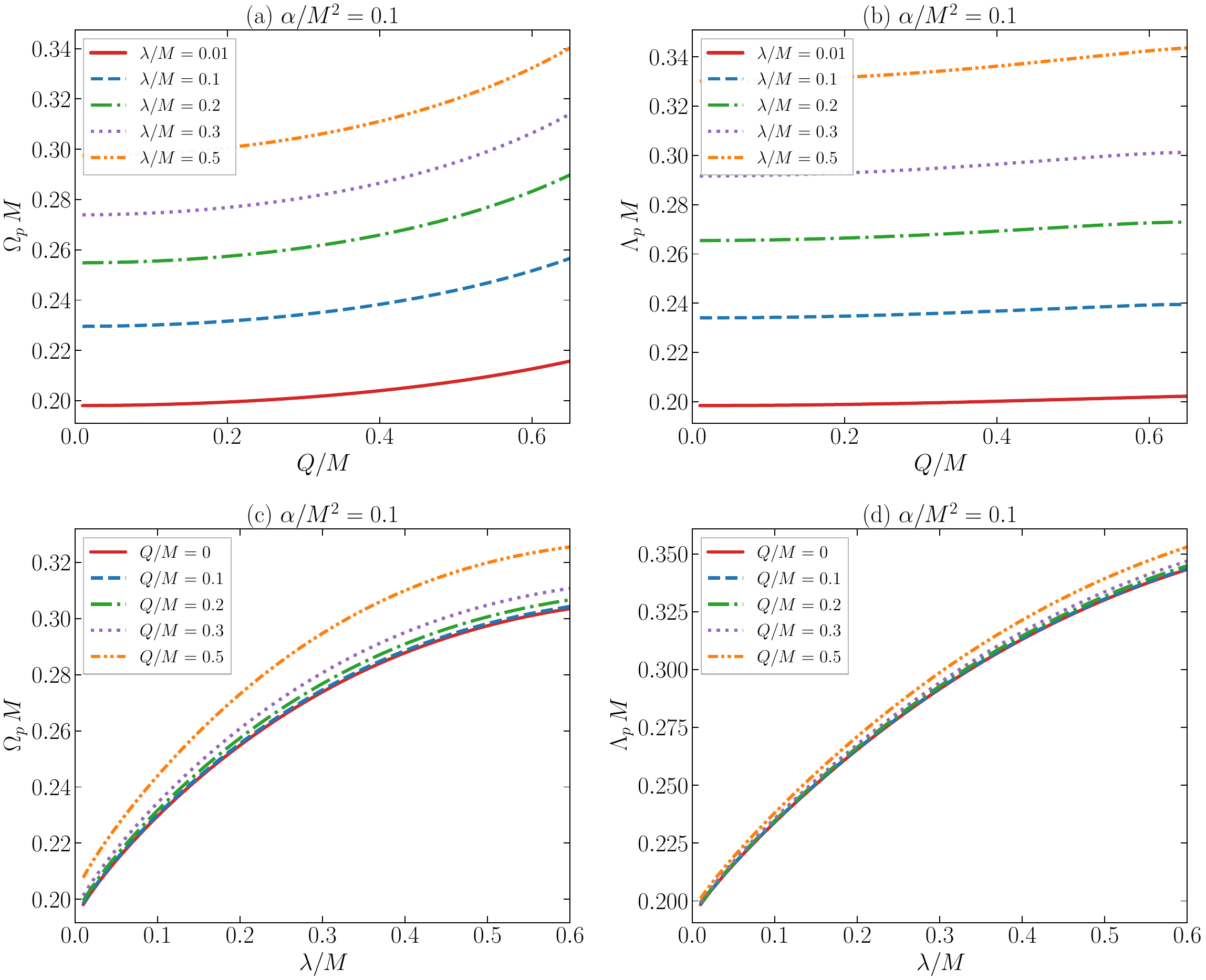}
    \caption{Eikonal QNM quantities at $\alpha/M^{2}=0.1$. (a)~Angular velocity $\Omega_p$ vs $Q/M$ for several $\lambda/M$. (b)~Lyapunov exponent $\Lambda_p$ vs $Q/M$. (c)~$\Omega_p$ vs $\lambda/M$ for several $Q/M$. (d)~$\Lambda_p$ vs $\lambda/M$.}
    \label{fig:qnm}
\end{figure*}

\begin{table*}[tbhp]
\centering
\begin{tabular}{|c|c|c|c|}
\hline
$\lambda/M$ & $M \omega_{0}$ & $M \omega_{1}$ & $M \omega_{2}$ \\
\hline
0.1 & $0.572360 - 0.188441\,i$ & $0.572360 - 0.565322\,i$ & $0.572360 - 0.942204\,i$ \\

0.2 & $0.634244 - 0.232359\,i$ & $0.634244 - 0.697078\,i$ & $0.634244 - 1.161797\,i$ \\

0.3 & $0.676107 - 0.264025\,i$ & $0.676107 - 0.792074\,i$ & $0.676107 - 1.320124\,i$ \\

0.4 & $0.704174 - 0.285732\,i$ & $0.704174 - 0.857195\,i$ & $0.704174 - 1.428658\,i$ \\

0.5 & $0.721971 - 0.299368\,i$ & $0.721971 - 0.898105\,i$ & $0.721971 - 1.496842\,i$ \\
\hline
\end{tabular}
\caption{QNMs frequencies $\omega_{n\ell}$ for $\ell=2$ with varying $\lambda$ and overtone numbers $n=0,1,2$. Here $Q/M=1,\,\alpha/M^2=0.10 \times 10^3$.}
\label{tab:5}
\hfill\\
\centering
\begin{tabular}{|c|c|c|c|}
\hline
$\lambda/M$ & $M \omega_{0}$ & $M \omega_{1}$ & $M \omega_{2}$ \\
\hline
0.1 & $0.554650 - 0.190253\,i$ & $0.554650 - 0.570759\,i$ & $0.554650 - 0.951266\,i$ \\

0.2 & $0.610875 - 0.231413\,i$ & $0.610875 - 0.694238\,i$ & $0.610875 - 1.157063\,i$ \\

0.3 & $0.649498 - 0.261402\,i$ & $0.649498 - 0.784207\,i$ & $0.649498 - 1.307012\,i$ \\

0.4 & $0.676057 - 0.282494\,i$ & $0.676057 - 0.847483\,i$ & $0.676057 - 1.412471\,i$ \\

0.5 & $0.693602 - 0.296363\,i$ & $0.693602 - 0.889088\,i$ & $0.693602 - 1.481813\,i$ \\
\hline
\end{tabular}
\caption{QNMs frequencies $\omega_{n\ell}$ for $\ell=2$ with varying $\lambda$ and overtone numbers $n=0,1,2$. Here $Q/M=1,\,\alpha/M^2=0.15 \times 10^3$.}
\label{tab:6}
\hfill\\
\centering
\begin{tabular}{|c|c|c|c|}
\hline
$\lambda/M$ & $M \omega_{0}$ & $M \omega_{1}$ & $M \omega_{2}$ \\
\hline
-0.5 & $0.275754 - 0.057440\,i$ & $0.275754 - 0.172320\,i$ & $0.275754 - 0.287200\,i$ \\

-0.4 & $0.291428 - 0.062020\,i$ & $0.291428 - 0.186060\,i$ & $0.291428 - 0.310099\,i$ \\

-0.3 & $0.311721 - 0.067907\,i$ & $0.311721 - 0.203721\,i$ & $0.311721 - 0.339535\,i$ \\

-0.2 & $0.339377 - 0.076136\,i$ & $0.339377 - 0.228409\,i$ & $0.339377 - 0.380682\,i$ \\

-0.1 & $0.380825 - 0.089623\,i$ & $0.380825 - 0.268870\,i$ & $0.380825 - 0.448116\,i$ \\
\hline
\end{tabular}
\caption{QNMs frequencies $\omega_{n\ell}$ for $\ell=2$ with varying $\lambda$ and overtone numbers $n=0,1,2$. Here $Q/M=1,\,\alpha/M^2=0.10 \times 10^3$.}
\label{tab:7}
\hfill\\
\begin{tabular}{|c|c|c|c|}
\hline
$\lambda/M$ & $M \omega_{0}$ & $M \omega_{1}$ & $M \omega_{2}$ \\
\hline
-0.5 & $0.275306 - 0.058110\,i$ & $0.275306 - 0.174330\,i$ & $0.275306 - 0.290550\,i$ \\

-0.4 & $0.290826 - 0.062891\,i$ & $0.290826 - 0.188673\,i$ & $0.290826 - 0.314456\,i$ \\

-0.3 & $0.310846 - 0.069116\,i$ & $0.310846 - 0.207349\,i$ & $0.310846 - 0.345582\,i$ \\

-0.2 & $0.337955 - 0.077947\,i$ & $0.337955 - 0.233840\,i$ & $0.337955 - 0.389734\,i$ \\

-0.1 & $0.378083 - 0.092542\,i$ & $0.378083 - 0.277627\,i$ & $0.378083 - 0.462712\,i$ \\
\hline
\end{tabular}
\caption{QNMs frequencies $\omega_{n\ell}$ for $\ell=2$ with varying $\lambda$ and overtone numbers $n=0,1,2$. Here $Q/M=1,\,\alpha/M^2=0.15 \times 10^3$.}
\label{tab:8}
\end{table*}

\section{Quasinormal Modes: Eikonal Limit}

In the eikonal regime $(\ell\gg 1)$, the quasinormal frequencies are determined by the properties of the unstable circular null geodesic. In this limit, they can be approximated as \cite{CardosoEtAl2009,SchutzWill1985,IyerWill1987}
\begin{equation}
\omega_{n\ell}\approx \ell\,\Omega_p-i\left(n+\tfrac{1}{2}\right)|\Lambda_p|,\label{qnm1-new}
\end{equation}
where $n$ is the overtone number, while $\Omega_p$ and $\Lambda_p$ are the angular velocity and Lyapunov exponent evaluated at the photon sphere radius $r_p$. These quantities are given by
\begin{align}
\Omega_p&=\sqrt{\frac{f(r_p)}{r_p^2}}=\frac{1}{R_{\rm sh}},\nonumber\\
\Lambda_p&=\Omega_p\sqrt{f(r_p)-r_p^2 f''(r_p)/2}=\frac{\sqrt{f(r_p)-r_p^2 f''(r_p)/2}}{R_{\rm sh}}.\label{qnm2-new}
\end{align}

Therefore, in the eikonal limit, the real part of the QNM frequency is controlled by the angular velocity of the photon sphere, whereas the imaginary part is determined by the instability timescale of the corresponding null orbit. This establishes a direct link between the shadow properties and the scalar quasinormal spectrum of the EH--PFDM black hole.

The behaviour of $\Omega_p$ and $\Lambda_p$ is presented in Fig.~\ref{fig:qnm}. Figures~\ref{fig:qnm}(a) and~s~\ref{fig:qnm}(b) show these quantities as functions of the charge for several values of~$\lambda$. At fixed $\lambda$, both $\Omega_p$ and $\Lambda_p$ increase with $Q$, indicating that more strongly charged black holes oscillate at higher real frequencies and decay faster. The PFDM parameter produces an even larger effect: at fixed $Q$ a larger $\lambda$ raises both quantities substantially, separating the curves into a well-resolved fan. In Figs.~\ref{fig:qnm}(c) and~\ref{fig:qnm}(d), the same quantities are plotted against $\lambda$ for different charges. Here, both $\Omega_p$ and $\Lambda_p$ grow monotonically with $\lambda$, with the charge providing only a moderate upward offset. This trend is fully consistent with the shadow analysis: a smaller photon sphere (driven by larger $\lambda$ or $Q$) corresponds to a higher angular velocity $\Omega_p=1/R_{\rm sh}$ and, simultaneously, to a steeper potential curvature that enhances the Lyapunov exponent. From a spectroscopic perspective, these results imply that the dominant imprint on the QNM spectrum comes from the dark-matter background, while the EH correction remains sub-dominant.

In Tables \ref{tab:5}--\ref{tab:8}, we presented the numerical values of the QMNs spectra for varying $\lambda$ for two values of EH parameter.

\begin{center}
    {\bf Damping Time and Quality Factor}
\end{center}

A convenient damping-time estimate is
\begin{equation}
    \tau_{n}=\frac{1}{|\text{Im}(\omega_{n \ell})|}=\frac{R_{\rm sh}}{(n+1/2)\,\sqrt{f(r_p)-r_p^2 f''(r_p)/2}}.
\end{equation}
As an additional analytic quantity, define the eikonal quality factor
\begin{equation}
    \mathcal{Q}_{n \ell}=\frac{\text{Re} (\omega_{n \ell})}{2|\text{Im} (\omega_{n \ell})|}=\frac{\ell\,\sqrt{f(r_p)-r_p^2 f''(r_p)/2}}{(2n+1)}.
\end{equation}

In the Schwarzschild limit (\(Q=0\) and \(\lambda \to 0\)), the photon sphere radius, shadow radius, damping time, and eikonal quality factor reduce to their corresponding standard Schwarzschild expressions \cite{Synge1966,Luminet1979,Mashhoon1985}:
\begin{align}
r_p=3M,\quad R_{\rm sh}=3\sqrt{3}M,\nonumber\\
\tau_{n}=\frac{3\sqrt{3}M }{n+\tfrac{1}{2}},\quad \mathcal{Q}_{n \ell}=\frac{\ell}{2 n+1}.
\end{align}

%Moreover, in the limit $\alpha \to 0$ and $\lambda \to 0$, the  aforementioned quantities simplifies as (provided $Q^2 \leq  9M^2/8$)
%\begin{align}
%r_p&=\frac{3M+\sqrt{9M^2-8Q^2}}{2},\nonumber\\
%R_{\rm sh}&=\frac{\frac{3M+\sqrt{9M^2-8Q^2}}{2}}{\sqrt{1-\frac{4M}{3M+\sqrt{9M^2-8Q^2}}+\frac{4Q^2}{(3M+\sqrt{9M^2-8Q^2})^2}}}\nonumber\\
%\tau_n&=\frac{R_{\rm sh}}{(n+\tfrac{1}{2})}\sqrt{\frac{9M^2+3M\sqrt{9M^2-8Q^2}-4Q^2}
%{9M^2+3M\sqrt{9M^2-8Q^2}-8Q^2}},\nonumber\\
%\mathcal{Q}_{n\ell}&=\frac{\ell}{2n+1}
%\sqrt{\frac{9M^2+3M\sqrt{9M^2-8Q^2}-8Q^2}{9M^2+3M\sqrt{9M^2-8Q^2}-4Q^2}}. 
%\end{align}

\section{Grey-body factors from the eikonal QNM correspondence}

For a fixed multipole \(\ell\), the GBF, also called the transmission probability, is defined by
\begin{equation}
\Gamma_\ell(\omega) \equiv |T_\ell(\omega)|^2, 
\qquad 
|R_\ell(\omega)|^2 + |T_\ell(\omega)|^2 = 1,
\end{equation}
where \(R_\ell\) and \(T_\ell\) are the reflection and transmission amplitudes in the one-dimensional scattering problem for the master wave equation. The associated scattering boundary conditions are
\begin{equation}
\Psi \to e^{-i\omega r_*} + R_\ell(\omega)e^{i\omega r_*} \quad (r_* \to +\infty),
\end{equation}
\begin{equation}
\Psi \to T_\ell(\omega)e^{-i\omega r_*} \quad (r_* \to -\infty),
\end{equation}
which imply flux conservation \(|R_\ell|^2 + |T_\ell|^2 = 1\) for real-frequency scattering in this static background.

In the eikonal regime, Ref.~\cite{KonoplyaZhidenko2024,KonoplyaZhidenko2025} gives the QNM--GBF correspondence in terms of the fundamental mode \(\omega_{0 \ell}\):
\begin{equation}
\Gamma_\ell(\omega) =\left[1 + \exp\left(2\pi \, \frac{\omega^2 - \mathrm{Re}(\omega_{0 \ell})^2}{4\,\mathrm{Re}(\omega_{0 \ell})\,\mathrm{Im}(\omega_{0 \ell})}\right)\right]^{-1}.
\end{equation}

Since $\text{Im}(\omega_{0 \ell})<0$ for damped modes, it is convenient to rewrite this as
\begin{equation}
\Gamma_\ell(\omega) =\left[1 +\exp\left\{-\pi\,\frac{R^2_{\rm sh}}{\ell\,\Theta_p}\left(\omega^2-\frac{\ell^2}{R^2_{\rm sh}}\right)\right\}\right]^{-1},\label{gbf}
\end{equation}
where $\Theta_p=\sqrt{f(r_p)-r_p^2\,f''(r_p)/2}$.

\begin{figure*}[tbhp]
    \centering
    \includegraphics[width=0.8\textwidth]{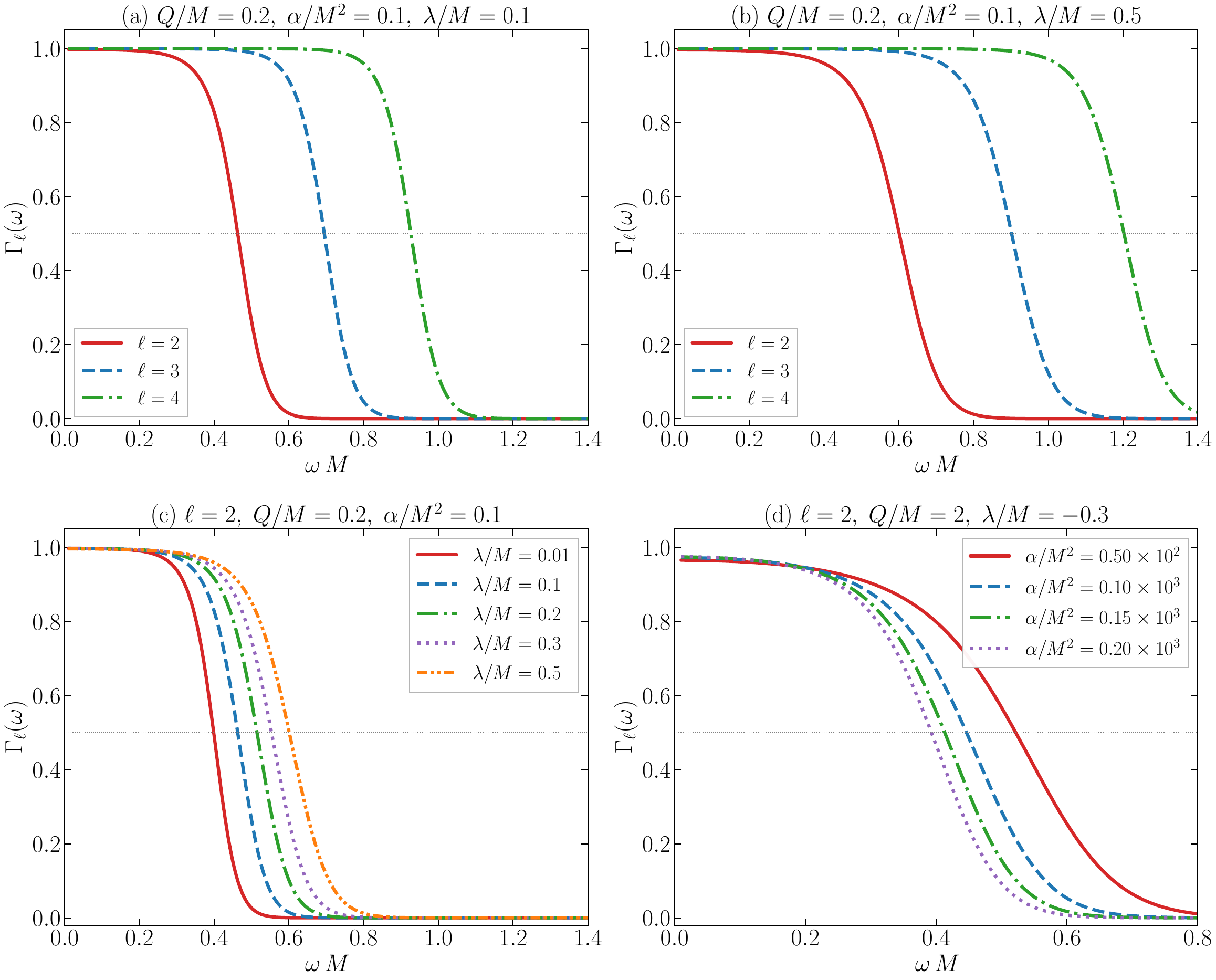}
    \caption{Eikonal grey-body factor $\Gamma_\ell(\omega)$ for the
    EH--PFDM black hole.
    (a)~Varying $\ell=2,\,3,\,4$ at $Q/M=0.2$, $\alpha/M^{2}=0.1$,
    $\lambda/M=0.1$.
    (b)~Varying $\ell=2,\,3,\,4$ at $Q/M=0.2$, $\alpha/M^{2}=0.1$,
    $\lambda/M=0.5$.
    (c)~Varying $\lambda/M$ at $\ell=2$, $Q/M=0.2$, $\alpha/M^{2}=0.1$.
    (d)~Varying $\alpha/M^{2}$ at $\ell=2$, $Q/M=2$, $\lambda/M=-0.3$.
    The horizontal dotted line marks $\Gamma_\ell=1/2$.}
    \label{fig:gbf}
\end{figure*}

% =====================================================================
%  GBF figure + discussion — to be inserted in Section VI
%  (Grey-Body Factors from the Eikonal QNM Correspondence),
%  right after Eq. (30).
% =====================================================================

The eikonal grey-body factor is plotted in Fig.~\ref{fig:gbf} for multipole
numbers $\ell=2,\,3,\,4$.  In all panels the transmission probability
exhibits the characteristic sigmoid profile predicted by Eq.~(\ref{gbf}):
$\Gamma_\ell\approx 1$ for low frequencies, where the wave passes above
the potential barrier almost unimpeded, and $\Gamma_\ell\to 0$ for
$\omega\gg\ell\,\Omega_p$, where the barrier reflects the wave back.  The
crossover $\Gamma_\ell=1/2$ occurs precisely at
$\omega=\ell\,\Omega_p=\ell/R_{\rm sh}$, linking the grey-body spectrum
directly to the shadow radius.

Figures~\ref{fig:gbf}(a) and~\ref{fig:gbf}(b) display the multipole
dependence of $\Gamma_\ell$ in two distinct dark-matter environments,
$\lambda/M=0.1$ and $\lambda/M=0.5$, respectively.  In both cases higher
$\ell$ shifts the transition to larger frequencies, reflecting the
proportionality $\mathrm{Re}(\omega_{0\ell})=\ell\,\Omega_p$.  Comparing
the two panels, the effect of the PFDM background is clearly visible: at
$\lambda/M=0.5$ all three curves are shifted to substantially higher
frequencies relative to the $\lambda/M=0.1$ case, and the transition
becomes steeper because the Lyapunov exponent $\Lambda_p$ grows with
$\lambda$.

Figure~\ref{fig:gbf}(c) isolates the $\lambda$-dependence at fixed
$\ell=2$, $Q/M=0.2$, and $\alpha/M^{2}=0.1$.  Raising $\lambda/M$ from
$0.01$ to $0.5$ shifts the crossover frequency by roughly a factor of
two while steepening the sigmoid, mirroring the shrinking of the shadow
radius.

In Fig.~\ref{fig:gbf}(d) the EH parameter $\alpha$ is varied at $\ell=2$, $Q/M=2$, and $\lambda/M=-0.3$, i.e., in the strong-charge regime where the $\alpha Q^{4}/(20\,r^{6})$ correction becomes appreciable.  Unlike the weak-charge panels, the curves are now clearly
separated: increasing $\alpha/M^{2}$ from $0.50\times10^{2}$ to $0.20\times10^{3}$ shifts the crossover from $\omega\,M\approx 0.52$ down to $\approx 0.39$, corresponding to a $\sim 25\%$ reduction in the critical frequency.  Physically, a larger $\alpha$ enlarges the photon sphere and the shadow radius, thereby lowering $\Omega_p=1/R_{\rm sh}$ and pushing the grey-body transition to softer frequencies.  The sigmoid also broadens, reflecting a smaller Lyapunov exponent at larger $r_p$.  These results demonstrate that the EH parameter leaves a significant imprint on the grey-body spectrum provided the electric charge is sufficiently large, complementing the PFDM-dominated behavior observed in Figs.~\ref{fig:gbf}(a)--(c).

%==========================================

\begin{figure*}[tbhp]
    \centering
    \includegraphics[width=0.8\linewidth]{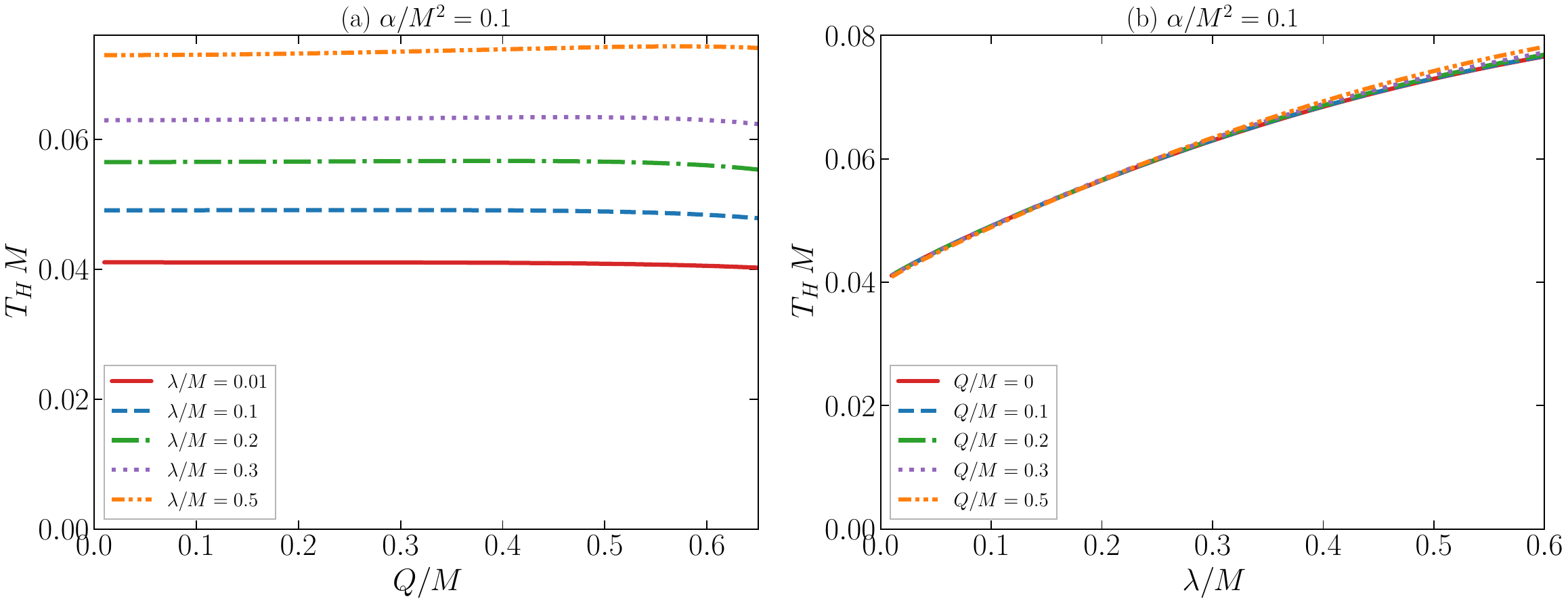}
    \caption{Hawking temperature $T_H\,M$ for the EH--PFDM black hole at $\alpha/M^{2}=0.1$. (a)~As a function of $Q/M$ for several values of $\lambda/M$. (b)~As a function of $\lambda/M$ for several values of $Q/M$.}
    \label{fig:temperature}
\end{figure*}

\begin{figure*}[tbhp]
    \centering
    \includegraphics[width=0.8\linewidth]{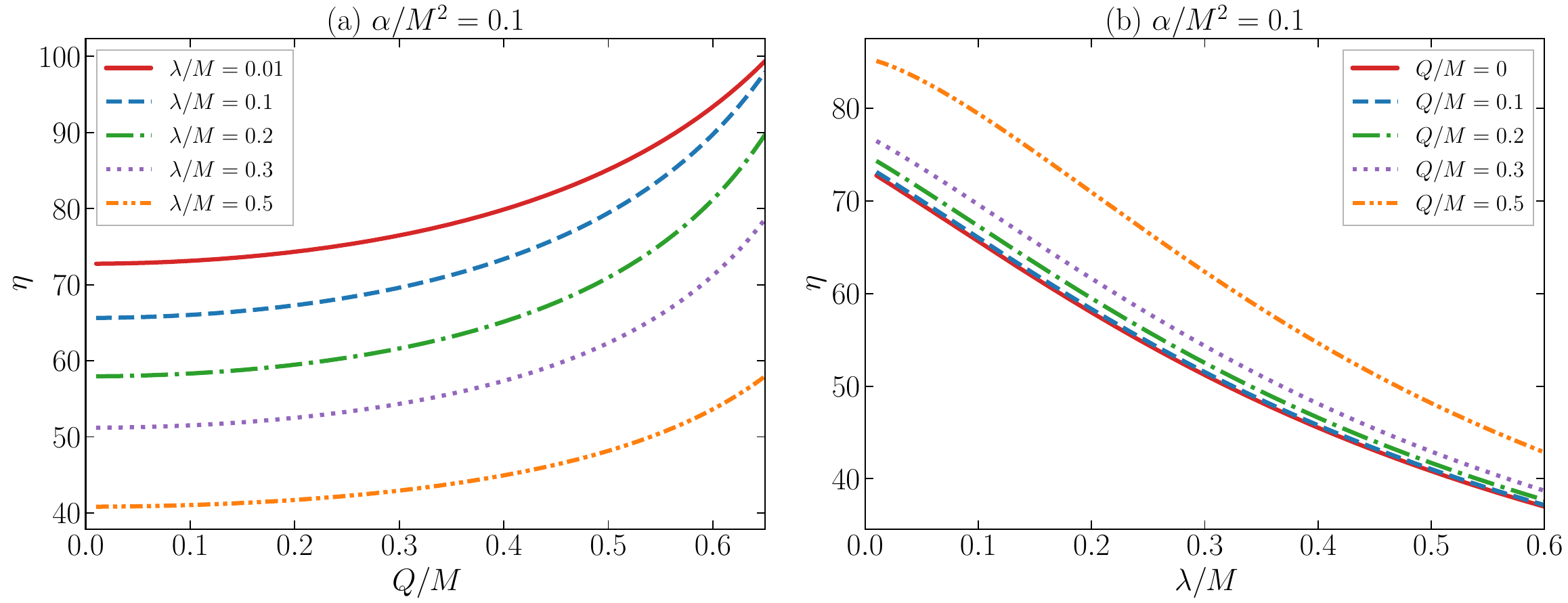}
    \caption{Dimensionless sparsity parameter $\eta$ of Hawking radiation at $\alpha/M^{2}=0.1$. (a)~As a function of $Q/M$ for several values of $\lambda/M$. (b)~As a function of $\lambda/M$ for several values of $Q/M$.}
    \label{fig:sparsity}
\end{figure*}

\section{Sparsity of Hawking Radiation}\label{sec:5}

We now examine the sparsity of Hawking radiation for the EH--PFDM black hole. The Hawking temperature, obtained from the surface gravity at the event horizon $r_h$, can be written as \cite{Bekenstein1973,Hawking1974,Hawking1975}
\begin{equation}
T_H=\frac{f'(r_h)}{4\pi}
=\frac{\Delta_h}{4\pi r_h},
\label{zz2-new}
\end{equation}
where we have introduced
\begin{equation}
\Delta_h\equiv
1-\frac{Q^2}{r_h^2}
+\frac{\alpha Q^4}{4r_h^6}
+\frac{\lambda}{r_h}.
\label{zz2b-new}
\end{equation}

Although Hawking radiation is thermal, the emission process is generally sparse, with quanta emitted at well-separated intervals. Following the standard definition \cite{Page1976a,GrayEtAl2015}, the dimensionless sparsity parameter is
\begin{equation}
\eta=\frac{\mathcal{C}}{\tilde g}\,
\frac{\lambda_t^2}{\mathcal{A}_{\rm eff}},
\qquad
\lambda_t=\frac{2\pi}{T_H},
\label{zz3-new}
\end{equation}
where $\mathcal{C}$ is a numerical constant, $\tilde g$ is the spin degeneracy factor, and the effective emitting area is
\begin{equation}
\mathcal{A}_{\rm eff}=\frac{27}{4}A_{\rm BH}=27\pi r_h^2.
\label{zz3b-new}
\end{equation}
\begin{figure*}[tbhp]
    \centering
    \includegraphics[width=\textwidth]{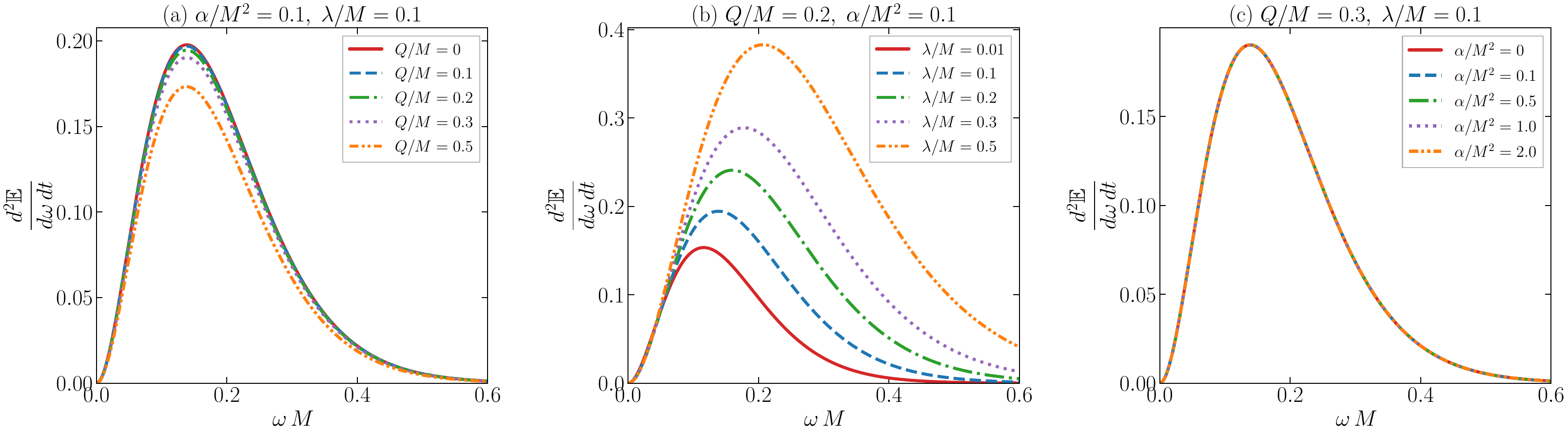}
    \caption{Spectral energy emission rate $d^{2}\mathbb{E}/(d\omega\,dt)$ for the EH--PFDM black hole. (a)~Varying $Q/M$ at $\alpha/M^{2}=0.1$ and $\lambda/M=0.1$. (b)~Varying $\lambda/M$ at $Q/M=0.2$ and $\alpha/M^{2}=0.1$. (c)~Varying $\alpha/M^{2}$ at $Q/M=0.3$ and $\lambda/M=0.1$.}
    \label{fig:emission}
\end{figure*}

Substituting Eq.~(\ref{zz2-new}) into Eq.~(\ref{zz3-new}), we obtain
\begin{equation}
\eta=\eta_{\rm Sch}\,\Delta_h^{-2},
\qquad
\eta_{\rm Sch}=\frac{64\pi^3}{27}.
\label{zz4-new}
\end{equation}

This compact result shows that the sparsity is fully controlled by the combination $\Delta_h$, which also determines the Hawking temperature.

The Hawking temperature is displayed in Fig.~\ref{fig:temperature}. Panel~(a) shows $T_H$ as a function of $Q$ for several values of~$\lambda$. At small $\lambda$ the temperature is only weakly sensitive to the charge, whereas at larger $\lambda$ the temperature is significantly enhanced and develops a mild decrease at high~$Q$ due to the growing Coulomb suppression in $\Delta_h$. Panel~(b) reveals a monotonic increase of $T_H$ with $\lambda$ at each fixed charge: the dark-matter contribution $\lambda/r_h$ adds positively to $\Delta_h$ and hence boosts the surface gravity. The charge dependence is comparatively weak and manifests mainly as a downward shift of the curves at large~$Q$. Taken together, these results show that the PFDM environment is the primary driver of the Hawking temperature in this system.

The sparsity parameter $\eta$ is shown in Fig.~\ref{fig:sparsity}. Since $\eta\propto\Delta_h^{-2}$, the trends are the inverse of those observed for $T_H$. In panel~(a) the sparsity increases with $Q$ at each fixed $\lambda$, reflecting the reduction of $\Delta_h$ by the Coulomb term $-Q^{2}/r_h^{2}$. Configurations with smaller $\lambda$ reach the highest sparsity values, exceeding $\eta\sim 95$ at $Q/M\simeq 0.65$ and $\lambda/M=0.01$. Panel~(b) shows that $\eta$ decreases monotonically with $\lambda$: the dark-matter contribution enhances the temperature and thereby compresses the thermal wavelength relative to the horizon area, making the radiation less sparse. In the range explored, $\eta$ remains well above unity for all parameter combinations, confirming that the Hawking cascade is genuinely sparse and cannot be approximated by a continuous blackbody emission.

\section{Energy Emission Rate}\label{sec:6}

We now analyze the energy emission rate associated with the EH-PFDM black hole. In the geometric-optics regime, the absorption cross-section approaches the limiting value
\begin{equation}
\sigma_{\rm lim}\approx \pi R_{\rm sh}^2=\pi\,\frac{r_p^2}{f(r_p)},\label{ee1-new}
\end{equation}
where $R_{\rm sh}$ is the shadow radius.

Within this approximation, the spectral energy emission rate is given by \cite{WeiLiu2013,DecaniniEtAl2011,Mashhoon1973,Misner1973}
\begin{equation}
\frac{d^2\mathbb{E}}{d\omega\,dt}=\frac{2\pi^2\sigma_{\rm lim}}{e^{\omega/T_H}-1}\,\omega^3,\label{ee2-new}
\end{equation}
where $\omega$ denotes the emitted frequency and $T_H$ is the Hawking temperature. Using Eqs.~(\ref{ee1-new}) and (\ref{zz2-new}), the emission rate can be written in the compact form
\begin{equation}
\frac{d^2\mathbb{E}}{d\omega\,dt}=\frac{2\pi^3 r_p^2\,\omega^3}{f(r_p)\left(e^{\omega/T_H}-1\right)}=\frac{2\pi^3 r_p^2\,\omega^3}
{f(r_p)\left(e^{4\pi\omega r_h/\Delta_h}-1\right)}.\label{ee3-new}
\end{equation}

This expression makes the underlying physics transparent. The shadow radius determines the effective absorption area, while the temperature is accounted for by the thermal factor. Hence, the EH parameter $\alpha$ and the PFDM parameter $\lambda$ affect the emission spectrum in two ways: indirectly through the photon sphere and shadow radius, and directly through the Hawking temperature.

The spectral emission rate is plotted in Fig.~\ref{fig:emission}. Panel~(a) examines the charge dependence at fixed $\alpha/M^{2}=0.1$ and $\lambda/M=0.1$. The peak amplitude decreases and shifts to slightly higher frequencies as $Q$ increases, consistent with the simultaneous reduction of the shadow area and the mild increase in temperature. In panel~(b) the PFDM parameter is varied at fixed $Q/M=0.2$. This produces the most dramatic effect: a larger $\lambda$ substantially enhances both the peak intensity and the peak frequency. The enhancement arises from two reinforcing mechanisms: the higher Hawking temperature broadens and blueshifts the thermal distribution, while the effective absorption cross section evolves more slowly, allowing the temperature effect to dominate. At $\lambda/M=0.5$ the peak intensity is roughly five times that of the $\lambda/M=0.01$ case. Panel~(c) varies the EH parameter at $Q/M=0.3$. As anticipated from the preceding analyses, the emission profiles are virtually indistinguishable, confirming that the leading-order QED correction has a negligible influence on the energy emission rate.

\section{Conclusions}\label{sec:7}

In this work, we presented a unified analysis of the optical, dynamical, and radiative properties of an Euler--Heisenberg black hole immersed in a perfect fluid dark matter background. More specifically, we studied the photon sphere and shadow, scalar perturbations and quasinormal modes in the eikonal limit, the grey-body factor through the eikonal QNM correspondence, the sparsity of Hawking radiation, and the corresponding spectral energy emission rate.

Our results show that the PFDM parameter and the electric charge substantially modify the relevant observables of the system. In particular, the photon sphere radius and the shadow radius are sensitive to the combined action of the charge and the dark-matter background, while the Euler--Heisenberg correction is generally subleading in the parameter range investigated. This same hierarchy extends to the eikonal QNM sector: through the geodesic correspondence, variations in the shadow radius are directly reflected in the angular velocity and instability timescale of the photon sphere, and hence in the real and imaginary parts of the quasinormal frequencies.

The scalar effective potential further supports this picture. Its dominant deformation is driven by the PFDM contribution, whereas the Euler--Heisenberg term produces only a comparatively small correction for most of the configurations considered. The same pattern is reflected in the grey-body factor: the PFDM background shifts the transition region more significantly, while the nonlinear electromagnetic correction becomes appreciable only in selected strong-charge regimes. This indicates that the radiative scattering properties are largely controlled by the surrounding matter distribution, with the Euler--Heisenberg parameter acting mainly as a refinement of the spectrum.

For the Hawking sector, we found that the temperature and sparsity are governed by the horizon combination entering $\Delta_h$. In the explored parameter domain, the PFDM parameter tends to enhance the Hawking temperature and reduce the sparsity, whereas the electric charge has the opposite tendency on the emission pattern. Nevertheless, the sparsity parameter remains well above unity throughout the parameter space analyzed, confirming that the Hawking cascade is genuinely sparse rather than continuous. The energy emission rate follows the same qualitative hierarchy: the dark-matter background produces the most visible changes in both the position and amplitude of the spectral peak, while the Euler--Heisenberg contribution remains weak in most cases.

Taken together, these results point to a clear phenomenological hierarchy within the approximations employed here. The PFDM environment provides the leading imprint on shadow observables, eikonal quasinormal modes, grey-body factors, and emission characteristics, whereas the Euler--Heisenberg nonlinear electrodynamics correction is typically secondary, although not always negligible in high-charge configurations. Therefore, within the present framework, shadow- and ringdown-related quantities appear more promising as probes of the surrounding dark-matter distribution than as precision diagnostics of the Euler--Heisenberg sector. A more complete assessment of the latter would likely require either stronger electromagnetic fields, a wider parameter range, or analyses beyond the eikonal and geometric-optics approximations.

\scriptsize 

\section*{Acknowledgments}

F. A. acknowledges the Inter University Centre for Astronomy and Astrophysics (IUCAA), Pune, India for granting visiting associateship. E. O. Silva acknowledges the support from Conselho Nacional de Desenvolvimento Cient\'{i}fico e Tecnol\'{o}gico (CNPq) (grants 306308/2022-3), Funda\c c\~ao de Amparo \`{a} Pesquisa e ao Desenvolvimento Cient\'{i}fico e Tecnol\'{o}gico do Maranh\~ao (FAPEMA) (grants UNIVERSAL-06395/22), and Coordena\c c\~ao de Aperfei\c coamento de Pessoal de N\'{i}vel Superior (CAPES) - Brazil (Code 001).


\scriptsize 
\begin{thebibliography}{59}%
	\makeatletter
	\providecommand \@ifxundefined [1]{%
		\@ifx{#1\undefined}
	}%
	\providecommand \@ifnum [1]{%
		\ifnum #1\expandafter \@firstoftwo
		\else \expandafter \@secondoftwo
		\fi
	}%
	\providecommand \@ifx [1]{%
		\ifx #1\expandafter \@firstoftwo
		\else \expandafter \@secondoftwo
		\fi
	}%
	\providecommand \natexlab [1]{#1}%
	\providecommand \enquote  [1]{``#1''}%
	\providecommand \bibnamefont  [1]{#1}%
	\providecommand \bibfnamefont [1]{#1}%
	\providecommand \citenamefont [1]{#1}%
	\providecommand \href@noop [0]{\@secondoftwo}%
	\providecommand \href [0]{\begingroup \@sanitize@url \@href}%
	\providecommand \@href[1]{\@@startlink{#1}\@@href}%
	\providecommand \@@href[1]{\endgroup#1\@@endlink}%
	\providecommand \@sanitize@url [0]{\catcode `\\12\catcode `\$12\catcode
		`\&12\catcode `\#12\catcode `\^12\catcode `\_12\catcode `\%12\relax}%
	\providecommand \@@startlink[1]{}%
	\providecommand \@@endlink[0]{}%
	\providecommand \url  [0]{\begingroup\@sanitize@url \@url }%
	\providecommand \@url [1]{\endgroup\@href {#1}{\urlprefix }}%
	\providecommand \urlprefix  [0]{URL }%
	\providecommand \Eprint [0]{\href }%
	\providecommand \doibase [0]{https://doi.org/}%
	\providecommand \selectlanguage [0]{\@gobble}%
	\providecommand \bibinfo  [0]{\@secondoftwo}%
	\providecommand \bibfield  [0]{\@secondoftwo}%
	\providecommand \translation [1]{[#1]}%
	\providecommand \BibitemOpen [0]{}%
	\providecommand \bibitemStop [0]{}%
	\providecommand \bibitemNoStop [0]{.\EOS\space}%
	\providecommand \EOS [0]{\spacefactor3000\relax}%
	\providecommand \BibitemShut  [1]{\csname bibitem#1\endcsname}%
	\let\auto@bib@innerbib\@empty
	%</preamble>
	\bibitem [{\citenamefont {Bekenstein}(1973)}]{Bekenstein1973}%
	\BibitemOpen
	\bibfield  {author} {\bibinfo {author} {\bibfnamefont {J.~D.}\ \bibnamefont
			{Bekenstein}},\ }\href {https://doi.org/10.1103/PhysRevD.7.2333} {\bibfield
		{journal} {\bibinfo  {journal} {Phys. Rev. D}\ }\textbf {\bibinfo {volume}
			{7}},\ \bibinfo {pages} {2333} (\bibinfo {year} {1973})}\BibitemShut
	{NoStop}%
	\bibitem [{\citenamefont {Hawking}(1975)}]{Hawking1975}%
	\BibitemOpen
	\bibfield  {author} {\bibinfo {author} {\bibfnamefont {S.~W.}\ \bibnamefont
			{Hawking}},\ }\href {https://doi.org/10.1007/BF02345020} {\bibfield
		{journal} {\bibinfo  {journal} {Commun. Math Phys.}\ }\textbf {\bibinfo
			{volume} {43}},\ \bibinfo {pages} {199} (\bibinfo {year} {1975})}\BibitemShut
	{NoStop}%
	\bibitem [{\citenamefont {Hawking}\ and\ \citenamefont
		{Page}(1983)}]{HawkingPage1983}%
	\BibitemOpen
	\bibfield  {author} {\bibinfo {author} {\bibfnamefont {S.~W.}\ \bibnamefont
			{Hawking}}\ and\ \bibinfo {author} {\bibfnamefont {D.~N.}\ \bibnamefont
			{Page}},\ }\href {https://doi.org/10.1007/BF01208266} {\bibfield  {journal}
		{\bibinfo  {journal} {Commun. Math Phys.}\ }\textbf {\bibinfo {volume}
			{87}},\ \bibinfo {pages} {577} (\bibinfo {year} {1983})}\BibitemShut
	{NoStop}%
	\bibitem [{\citenamefont {Regge}\ and\ \citenamefont
		{Wheeler}(1957)}]{ReggeWheeler1957}%
	\BibitemOpen
	\bibfield  {author} {\bibinfo {author} {\bibfnamefont {T.}~\bibnamefont
			{Regge}}\ and\ \bibinfo {author} {\bibfnamefont {J.~A.}\ \bibnamefont
			{Wheeler}},\ }\href {https://doi.org/10.1103/PhysRev.108.1063} {\bibfield
		{journal} {\bibinfo  {journal} {Phys. Rev.}\ }\textbf {\bibinfo {volume}
			{108}},\ \bibinfo {pages} {1063} (\bibinfo {year} {1957})}\BibitemShut
	{NoStop}%
	\bibitem [{\citenamefont {Zerilli}(1970)}]{Zerilli1970}%
	\BibitemOpen
	\bibfield  {author} {\bibinfo {author} {\bibfnamefont {F.~J.}\ \bibnamefont
			{Zerilli}},\ }\href {https://doi.org/10.1103/PhysRevLett.24.737} {\bibfield
		{journal} {\bibinfo  {journal} {Phys. Rev. Lett.}\ }\textbf {\bibinfo
			{volume} {24}},\ \bibinfo {pages} {737} (\bibinfo {year} {1970})}\BibitemShut
	{NoStop}%
	\bibitem [{\citenamefont {Vishveshwara}(1970)}]{Vishveshwara1970}%
	\BibitemOpen
	\bibfield  {author} {\bibinfo {author} {\bibfnamefont {C.~V.}\ \bibnamefont
			{Vishveshwara}},\ }\href {https://doi.org/10.1038/227936a0} {\bibfield
		{journal} {\bibinfo  {journal} {Nature}\ }\textbf {\bibinfo {volume} {227}},\
		\bibinfo {pages} {936} (\bibinfo {year} {1970})}\BibitemShut {NoStop}%
	\bibitem [{\citenamefont {Kokkotas}\ and\ \citenamefont
		{Schmidt}(1999)}]{KokkotasSchmidt1999}%
	\BibitemOpen
	\bibfield  {author} {\bibinfo {author} {\bibfnamefont {K.~D.}\ \bibnamefont
			{Kokkotas}}\ and\ \bibinfo {author} {\bibfnamefont {B.~G.}\ \bibnamefont
			{Schmidt}},\ }\href {https://doi.org/10.12942/lrr-1999-2} {\bibfield
		{journal} {\bibinfo  {journal} {Living Rev. Relativ.}\ }\textbf {\bibinfo
			{volume} {2}},\ \bibinfo {pages} {2} (\bibinfo {year} {1999})}\BibitemShut
	{NoStop}%
	\bibitem [{\citenamefont {Berti}\ \emph {et~al.}(2009)\citenamefont {Berti},
		\citenamefont {Cardoso},\ and\ \citenamefont
		{Starinets}}]{BertiCardosoStarinets2009}%
	\BibitemOpen
	\bibfield  {author} {\bibinfo {author} {\bibfnamefont {E.}~\bibnamefont
			{Berti}}, \bibinfo {author} {\bibfnamefont {V.}~\bibnamefont {Cardoso}},\
		and\ \bibinfo {author} {\bibfnamefont {A.~O.}\ \bibnamefont {Starinets}},\
	}\href {https://doi.org/10.1088/0264-9381/26/16/163001} {\bibfield  {journal}
		{\bibinfo  {journal} {Class. Quantum Grav.}\ }\textbf {\bibinfo {volume}
			{26}},\ \bibinfo {pages} {163001} (\bibinfo {year} {2009})}\BibitemShut
	{NoStop}%
	\bibitem [{\citenamefont {Heisenberg}\ and\ \citenamefont
		{Euler}(1936{\natexlab{a}})}]{HeisenbergEuler1936}%
	\BibitemOpen
	\bibfield  {author} {\bibinfo {author} {\bibfnamefont {W.}~\bibnamefont
			{Heisenberg}}\ and\ \bibinfo {author} {\bibfnamefont {H.}~\bibnamefont
			{Euler}},\ }\href {https://doi.org/10.1007/BF01343663} {\bibfield  {journal}
		{\bibinfo  {journal} {Zeits. f{\"u}r Phys. (Berlin)}\ }\textbf {\bibinfo
			{volume} {98}},\ \bibinfo {pages} {714} (\bibinfo {year}
		{1936}{\natexlab{a}})}\BibitemShut {NoStop}%
	\bibitem [{\citenamefont {Plebanski}(1970)}]{Plebanski1970}%
	\BibitemOpen
	\bibfield  {author} {\bibinfo {author} {\bibfnamefont {J.~F.}\ \bibnamefont
			{Plebanski}},\ }\href@noop {} {\emph {\bibinfo {title} {Lectures on Nonlinear
				Electrodynamics}}}\ (\bibinfo  {publisher} {Nordita},\ \bibinfo {address}
	{Copenhagen},\ \bibinfo {year} {1970})\BibitemShut {NoStop}%
	\bibitem [{\citenamefont {Salazar}\ \emph
		{et~al.}(1987{\natexlab{a}})\citenamefont {Salazar}, \citenamefont
		{Garc{\'\i}a},\ and\ \citenamefont
		{Pleba{\'n}ski}}]{SalazarGarciaPlebanski1987}%
	\BibitemOpen
	\bibfield  {author} {\bibinfo {author} {\bibfnamefont {H.}~\bibnamefont
			{Salazar}}, \bibinfo {author} {\bibfnamefont {A.}~\bibnamefont
			{Garc{\'\i}a}},\ and\ \bibinfo {author} {\bibfnamefont {J.~F.}\ \bibnamefont
			{Pleba{\'n}ski}},\ }\href {https://doi.org/10.1063/1.527833} {\bibfield
		{journal} {\bibinfo  {journal} {J. Math. Phys.}\ }\textbf {\bibinfo {volume}
			{28}},\ \bibinfo {pages} {2171} (\bibinfo {year}
		{1987}{\natexlab{a}})}\BibitemShut {NoStop}%
	\bibitem [{\citenamefont {Yajima}\ and\ \citenamefont
		{Tamaki}(2001{\natexlab{a}})}]{YajimaTamaki2001}%
	\BibitemOpen
	\bibfield  {author} {\bibinfo {author} {\bibfnamefont {H.}~\bibnamefont
			{Yajima}}\ and\ \bibinfo {author} {\bibfnamefont {T.}~\bibnamefont
			{Tamaki}},\ }\href {https://doi.org/10.1103/PhysRevD.63.064007} {\bibfield
		{journal} {\bibinfo  {journal} {Phys. Rev. D}\ }\textbf {\bibinfo {volume}
			{63}},\ \bibinfo {pages} {064007} (\bibinfo {year}
		{2001}{\natexlab{a}})}\BibitemShut {NoStop}%
	\bibitem [{\citenamefont {Breton}\ and\ \citenamefont
		{L{\'o}pez}(2021)}]{BretonLopez2021}%
	\BibitemOpen
	\bibfield  {author} {\bibinfo {author} {\bibfnamefont {N.}~\bibnamefont
			{Breton}}\ and\ \bibinfo {author} {\bibfnamefont {L.~A.}\ \bibnamefont
			{L{\'o}pez}},\ }\href {https://doi.org/10.1103/PhysRevD.104.024064}
	{\bibfield  {journal} {\bibinfo  {journal} {Phys. Rev. D}\ }\textbf {\bibinfo
			{volume} {104}},\ \bibinfo {pages} {024064} (\bibinfo {year}
		{2021})}\BibitemShut {NoStop}%
	\bibitem [{\citenamefont {Rahaman}\ \emph {et~al.}(2010)\citenamefont
		{Rahaman}, \citenamefont {Nandi}, \citenamefont {Bhadra}, \citenamefont
		{Kalam},\ and\ \citenamefont {Chakraborty}}]{RahamanEtAl2010}%
	\BibitemOpen
	\bibfield  {author} {\bibinfo {author} {\bibfnamefont {F.}~\bibnamefont
			{Rahaman}}, \bibinfo {author} {\bibfnamefont {K.~K.}\ \bibnamefont {Nandi}},
		\bibinfo {author} {\bibfnamefont {A.}~\bibnamefont {Bhadra}}, \bibinfo
		{author} {\bibfnamefont {M.}~\bibnamefont {Kalam}},\ and\ \bibinfo {author}
		{\bibfnamefont {K.}~\bibnamefont {Chakraborty}},\ }\href
	{https://doi.org/10.1016/j.physletb.2010.10.022} {\bibfield  {journal}
		{\bibinfo  {journal} {Phys. Lett. B}\ }\textbf {\bibinfo {volume} {694}},\
		\bibinfo {pages} {10} (\bibinfo {year} {2010})}\BibitemShut {NoStop}%
	\bibitem [{\citenamefont {Li}\ and\ \citenamefont
		{Yang}(2012{\natexlab{a}})}]{LiYang2012}%
	\BibitemOpen
	\bibfield  {author} {\bibinfo {author} {\bibfnamefont {M.-H.}\ \bibnamefont
			{Li}}\ and\ \bibinfo {author} {\bibfnamefont {K.-C.}\ \bibnamefont {Yang}},\
	}\href {https://doi.org/10.1103/PhysRevD.86.123015} {\bibfield  {journal}
		{\bibinfo  {journal} {Phys. Rev. D}\ }\textbf {\bibinfo {volume} {86}},\
		\bibinfo {pages} {123015} (\bibinfo {year} {2012}{\natexlab{a}})}\BibitemShut
	{NoStop}%
	\bibitem [{\citenamefont {Haroon}\ \emph {et~al.}(2019)\citenamefont {Haroon},
		\citenamefont {Jamil}, \citenamefont {Jusufi}, \citenamefont {Lin},\ and\
		\citenamefont {Mann}}]{HaroonEtAl2019}%
	\BibitemOpen
	\bibfield  {author} {\bibinfo {author} {\bibfnamefont {S.}~\bibnamefont
			{Haroon}}, \bibinfo {author} {\bibfnamefont {M.}~\bibnamefont {Jamil}},
		\bibinfo {author} {\bibfnamefont {K.}~\bibnamefont {Jusufi}}, \bibinfo
		{author} {\bibfnamefont {K.}~\bibnamefont {Lin}},\ and\ \bibinfo {author}
		{\bibfnamefont {R.~B.}\ \bibnamefont {Mann}},\ }\href
	{https://doi.org/10.1103/PhysRevD.99.044015} {\bibfield  {journal} {\bibinfo
			{journal} {Phys. Rev. D}\ }\textbf {\bibinfo {volume} {99}},\ \bibinfo
		{pages} {044015} (\bibinfo {year} {2019})}\BibitemShut {NoStop}%
	\bibitem [{\citenamefont {Ahmed}\ \emph {et~al.}(2025)\citenamefont {Ahmed},
		\citenamefont {Al-Badawi},\ and\ \citenamefont {Sakalli}}]{Ahmed2025}%
	\BibitemOpen
	\bibfield  {author} {\bibinfo {author} {\bibfnamefont {F.}~\bibnamefont
			{Ahmed}}, \bibinfo {author} {\bibfnamefont {A.}~\bibnamefont {Al-Badawi}},\
		and\ \bibinfo {author} {\bibfnamefont {I.}~\bibnamefont {Sakalli}},\ }\href
	{https://doi.org/10.48550/arXiv.2509.12264} {} (\bibinfo {year} {2025}),\
	\Eprint {https://arxiv.org/abs/2509.12264} {arXiv:2509.12264 [gr-qc]}
	\BibitemShut {NoStop}%
	\bibitem [{\citenamefont {Shahzad}\ \emph {et~al.}(2025)\citenamefont
		{Shahzad}, \citenamefont {Abbas}, \citenamefont {Zhu}, \citenamefont {Ali},
		\citenamefont {Ashraf},\ and\ \citenamefont {Al-Kahtani}}]{Shahzad2025}%
	\BibitemOpen
	\bibfield  {author} {\bibinfo {author} {\bibfnamefont {M.~R.}\ \bibnamefont
			{Shahzad}}, \bibinfo {author} {\bibfnamefont {G.}~\bibnamefont {Abbas}},
		\bibinfo {author} {\bibfnamefont {T.}~\bibnamefont {Zhu}}, \bibinfo {author}
		{\bibfnamefont {R.~H.}\ \bibnamefont {Ali}}, \bibinfo {author} {\bibfnamefont
			{A.}~\bibnamefont {Ashraf}},\ and\ \bibinfo {author} {\bibfnamefont {B.~S.}\
			\bibnamefont {Al-Kahtani}},\ }\href
	{https://doi.org/10.1140/epjc/s10052-025-13876-w} {\bibfield  {journal}
		{\bibinfo  {journal} {Eur. Phys. J C}\ }\textbf {\bibinfo {volume} {85}},\
		\bibinfo {pages} {164} (\bibinfo {year} {2025})}\BibitemShut {NoStop}%
	\bibitem [{\citenamefont {Ma}\ \emph {et~al.}(2024)\citenamefont {Ma},
		\citenamefont {Wang}, \citenamefont {Deng},\ and\ \citenamefont
		{Hu}}]{Ma2024}%
	\BibitemOpen
	\bibfield  {author} {\bibinfo {author} {\bibfnamefont {S.~J.}\ \bibnamefont
			{Ma}}, \bibinfo {author} {\bibfnamefont {R.~B.}\ \bibnamefont {Wang}},
		\bibinfo {author} {\bibfnamefont {J.~B.}\ \bibnamefont {Deng}},\ and\
		\bibinfo {author} {\bibfnamefont {X.-R.}\ \bibnamefont {Hu}},\ }\href
	{https://doi.org/10.1140/epjc/s10052-024-12914-3} {\bibfield  {journal}
		{\bibinfo  {journal} {Eur. Phys. J C}\ }\textbf {\bibinfo {volume} {84}},\
		\bibinfo {pages} {595} (\bibinfo {year} {2024})}\BibitemShut {NoStop}%
	\bibitem [{\citenamefont {Ahmed}\ \emph
		{et~al.}(2026{\natexlab{a}})\citenamefont {Ahmed}, \citenamefont
		{Al-Badawi},\ and\ \citenamefont {Sakalli}}]{Ahmed2026ChargedBardeen}%
	\BibitemOpen
	\bibfield  {author} {\bibinfo {author} {\bibfnamefont {F.}~\bibnamefont
			{Ahmed}}, \bibinfo {author} {\bibfnamefont {A.}~\bibnamefont {Al-Badawi}},\
		and\ \bibinfo {author} {\bibfnamefont {I.}~\bibnamefont {Sakalli}},\ }\href
	{https://arxiv.org/abs/2602.02586} {} (\bibinfo {year}
	{2026}{\natexlab{a}}),\ \Eprint {https://arxiv.org/abs/2602.02586}
	{arXiv:2602.02586 [gr-qc]} \BibitemShut {NoStop}%
	\bibitem [{\citenamefont {Ahmed}\ \emph
		{et~al.}(2026{\natexlab{b}})\citenamefont {Ahmed}, \citenamefont
		{Al-Badawi},\ and\ \citenamefont {Sakalli}}]{Ahmed2026DarkMatterStringCloud}%
	\BibitemOpen
	\bibfield  {author} {\bibinfo {author} {\bibfnamefont {F.}~\bibnamefont
			{Ahmed}}, \bibinfo {author} {\bibfnamefont {A.}~\bibnamefont {Al-Badawi}},\
		and\ \bibinfo {author} {\bibfnamefont {I.}~\bibnamefont {Sakalli}},\ }\href
	{https://arxiv.org/abs/2602.02621} {} (\bibinfo {year}
	{2026}{\natexlab{b}}),\ \Eprint {https://arxiv.org/abs/2602.02621}
	{arXiv:2602.02621 [gr-qc]} \BibitemShut {NoStop}%
	\bibitem [{\citenamefont {Al-Badawi}\ \emph {et~al.}(2026)\citenamefont
		{Al-Badawi}, \citenamefont {Ahmed},\ and\ \citenamefont
		{Sakalli}}]{AlBadawi2026BlackHoleStudy}%
	\BibitemOpen
	\bibfield  {author} {\bibinfo {author} {\bibfnamefont {A.}~\bibnamefont
			{Al-Badawi}}, \bibinfo {author} {\bibfnamefont {F.}~\bibnamefont {Ahmed}},\
		and\ \bibinfo {author} {\bibfnamefont {I.}~\bibnamefont {Sakalli}},\ }\href
	{https://arxiv.org/abs/2603.02276} {} (\bibinfo {year} {2026}),\ \Eprint
	{https://arxiv.org/abs/2603.02276} {arXiv:2603.02276 [gr-qc]} \BibitemShut
	{NoStop}%
	\bibitem [{\citenamefont {Ahmed}\ \emph
		{et~al.}(2026{\natexlab{c}})\citenamefont {Ahmed}, \citenamefont
		{Al-Badawi},\ and\ \citenamefont {Silva}}]{Ahmed2026ChargedBHNonlinear}%
	\BibitemOpen
	\bibfield  {author} {\bibinfo {author} {\bibfnamefont {F.}~\bibnamefont
			{Ahmed}}, \bibinfo {author} {\bibfnamefont {A.}~\bibnamefont {Al-Badawi}},\
		and\ \bibinfo {author} {\bibfnamefont {E.~O.}\ \bibnamefont {Silva}},\ }\href
	{https://arxiv.org/abs/2602.07806} {} (\bibinfo {year}
	{2026}{\natexlab{c}}),\ \Eprint {https://arxiv.org/abs/2602.07806}
	{arXiv:2602.07806 [gr-qc]} \BibitemShut {NoStop}%
	\bibitem [{\citenamefont {Synge}(1966)}]{Synge1966}%
	\BibitemOpen
	\bibfield  {author} {\bibinfo {author} {\bibfnamefont {J.~L.}\ \bibnamefont
			{Synge}},\ }\href {https://doi.org/10.1093/mnras/131.3.463} {\bibfield
		{journal} {\bibinfo  {journal} {MNRAS}\ }\textbf {\bibinfo {volume} {131}},\
		\bibinfo {pages} {463} (\bibinfo {year} {1966})}\BibitemShut {NoStop}%
	\bibitem [{\citenamefont {Cunningham}\ and\ \citenamefont
		{Bardeen}(1973)}]{CunninghamBardeen1973}%
	\BibitemOpen
	\bibfield  {author} {\bibinfo {author} {\bibfnamefont {C.~T.}\ \bibnamefont
			{Cunningham}}\ and\ \bibinfo {author} {\bibfnamefont {J.~M.}\ \bibnamefont
			{Bardeen}},\ }\href {https://doi.org/10.1086/152223} {\bibfield  {journal}
		{\bibinfo  {journal} {Astrophys. J}\ }\textbf {\bibinfo {volume} {183}},\
		\bibinfo {pages} {237} (\bibinfo {year} {1973})}\BibitemShut {NoStop}%
	\bibitem [{\citenamefont {Luminet}(1979)}]{Luminet1979}%
	\BibitemOpen
	\bibfield  {author} {\bibinfo {author} {\bibfnamefont {J.-P.}\ \bibnamefont
			{Luminet}},\ }\href@noop {} {\bibfield  {journal} {\bibinfo  {journal}
			{Astron. \& Astrophys.}\ }\textbf {\bibinfo {volume} {75}},\ \bibinfo {pages}
		{228} (\bibinfo {year} {1979})}\BibitemShut {NoStop}%
	\bibitem [{\citenamefont {Hioki}\ and\ \citenamefont
		{Maeda}(2009)}]{HiokiMaeda2009}%
	\BibitemOpen
	\bibfield  {author} {\bibinfo {author} {\bibfnamefont {K.}~\bibnamefont
			{Hioki}}\ and\ \bibinfo {author} {\bibfnamefont {K.-i.}\ \bibnamefont
			{Maeda}},\ }\href {https://doi.org/10.1103/PhysRevD.80.024042} {\bibfield
		{journal} {\bibinfo  {journal} {Phys. Rev. D}\ }\textbf {\bibinfo {volume}
			{80}},\ \bibinfo {pages} {024042} (\bibinfo {year} {2009})}\BibitemShut
	{NoStop}%
	\bibitem [{\citenamefont {Perlick}\ and\ \citenamefont
		{Tsupko}(2022)}]{PerlickTsupko2022}%
	\BibitemOpen
	\bibfield  {author} {\bibinfo {author} {\bibfnamefont {V.}~\bibnamefont
			{Perlick}}\ and\ \bibinfo {author} {\bibfnamefont {O.~Y.}\ \bibnamefont
			{Tsupko}},\ }\href {https://doi.org/10.1016/j.physrep.2021.10.004} {\bibfield
		{journal} {\bibinfo  {journal} {Phys. Rep.}\ }\textbf {\bibinfo {volume}
			{947}},\ \bibinfo {pages} {1} (\bibinfo {year} {2022})}\BibitemShut {NoStop}%
	\bibitem [{\citenamefont {Akiyama}\ \emph
		{et~al.}(2019{\natexlab{a}})\citenamefont {Akiyama} \emph
		{et~al.}}]{EHTM87I2019}%
	\BibitemOpen
	\bibfield  {author} {\bibinfo {author} {\bibfnamefont {K.}~\bibnamefont
			{Akiyama}} \emph {et~al.},\ }\href {https://doi.org/10.3847/2041-8213/ab0ec7}
	{\bibfield  {journal} {\bibinfo  {journal} {Astrophys. J Lett.}\ }\textbf
		{\bibinfo {volume} {875}},\ \bibinfo {pages} {L1} (\bibinfo {year}
		{2019}{\natexlab{a}})}\BibitemShut {NoStop}%
	\bibitem [{\citenamefont {Akiyama}\ \emph
		{et~al.}(2019{\natexlab{b}})\citenamefont {Akiyama} \emph
		{et~al.}}]{EHTM87VI2019}%
	\BibitemOpen
	\bibfield  {author} {\bibinfo {author} {\bibfnamefont {K.}~\bibnamefont
			{Akiyama}} \emph {et~al.},\ }\href {https://doi.org/10.3847/2041-8213/ab1141}
	{\bibfield  {journal} {\bibinfo  {journal} {Astrophys. J Lett.}\ }\textbf
		{\bibinfo {volume} {875}},\ \bibinfo {pages} {L6} (\bibinfo {year}
		{2019}{\natexlab{b}})}\BibitemShut {NoStop}%
	\bibitem [{\citenamefont {Akiyama}\ \emph
		{et~al.}(2022{\natexlab{a}})\citenamefont {Akiyama} \emph
		{et~al.}}]{EHTSgrAI2022}%
	\BibitemOpen
	\bibfield  {author} {\bibinfo {author} {\bibfnamefont {K.}~\bibnamefont
			{Akiyama}} \emph {et~al.},\ }\href {https://doi.org/10.3847/2041-8213/ac6674}
	{\bibfield  {journal} {\bibinfo  {journal} {Astrophys. J Lett.}\ }\textbf
		{\bibinfo {volume} {930}},\ \bibinfo {pages} {L12} (\bibinfo {year}
		{2022}{\natexlab{a}})}\BibitemShut {NoStop}%
	\bibitem [{\citenamefont {Akiyama}\ \emph
		{et~al.}(2022{\natexlab{b}})\citenamefont {Akiyama} \emph
		{et~al.}}]{EHTSgrAVI2022}%
	\BibitemOpen
	\bibfield  {author} {\bibinfo {author} {\bibfnamefont {K.}~\bibnamefont
			{Akiyama}} \emph {et~al.},\ }\href@noop {} {\bibfield  {journal} {\bibinfo
			{journal} {Astrophys. J Lett.}\ }\textbf {\bibinfo {volume} {930}},\ \bibinfo
		{pages} {L17} (\bibinfo {year} {2022}{\natexlab{b}})}\BibitemShut {NoStop}%
	\bibitem [{\citenamefont {Schutz}\ and\ \citenamefont
		{Will}(1985)}]{SchutzWill1985}%
	\BibitemOpen
	\bibfield  {author} {\bibinfo {author} {\bibfnamefont {B.~F.}\ \bibnamefont
			{Schutz}}\ and\ \bibinfo {author} {\bibfnamefont {C.~M.}\ \bibnamefont
			{Will}},\ }\href {https://doi.org/10.1086/184453} {\bibfield  {journal}
		{\bibinfo  {journal} {Astrophys. J Lett.}\ }\textbf {\bibinfo {volume}
			{291}},\ \bibinfo {pages} {L33} (\bibinfo {year} {1985})}\BibitemShut
	{NoStop}%
	\bibitem [{\citenamefont {Iyer}\ and\ \citenamefont
		{Will}(1987)}]{IyerWill1987}%
	\BibitemOpen
	\bibfield  {author} {\bibinfo {author} {\bibfnamefont {S.}~\bibnamefont
			{Iyer}}\ and\ \bibinfo {author} {\bibfnamefont {C.~M.}\ \bibnamefont
			{Will}},\ }\href {https://doi.org/10.1103/PhysRevD.35.3621} {\bibfield
		{journal} {\bibinfo  {journal} {Phys. Rev. D}\ }\textbf {\bibinfo {volume}
			{35}},\ \bibinfo {pages} {3621} (\bibinfo {year} {1987})}\BibitemShut
	{NoStop}%
	\bibitem [{\citenamefont {Cardoso}\ \emph {et~al.}(2009)\citenamefont
		{Cardoso}, \citenamefont {Miranda}, \citenamefont {Berti}, \citenamefont
		{Witek},\ and\ \citenamefont {Zanchin}}]{CardosoEtAl2009}%
	\BibitemOpen
	\bibfield  {author} {\bibinfo {author} {\bibfnamefont {V.}~\bibnamefont
			{Cardoso}}, \bibinfo {author} {\bibfnamefont {A.~S.}\ \bibnamefont
			{Miranda}}, \bibinfo {author} {\bibfnamefont {E.}~\bibnamefont {Berti}},
		\bibinfo {author} {\bibfnamefont {H.}~\bibnamefont {Witek}},\ and\ \bibinfo
		{author} {\bibfnamefont {V.~T.}\ \bibnamefont {Zanchin}},\ }\href
	{https://doi.org/10.1103/PhysRevD.79.064016} {\bibfield  {journal} {\bibinfo
			{journal} {Phys. Rev. D}\ }\textbf {\bibinfo {volume} {79}},\ \bibinfo
		{pages} {064016} (\bibinfo {year} {2009})}\BibitemShut {NoStop}%
	\bibitem [{\citenamefont {Page}(1976{\natexlab{a}})}]{Page1976a}%
	\BibitemOpen
	\bibfield  {author} {\bibinfo {author} {\bibfnamefont {D.~N.}\ \bibnamefont
			{Page}},\ }\href {https://doi.org/10.1103/PhysRevD.13.198} {\bibfield
		{journal} {\bibinfo  {journal} {Phys. Rev. D}\ }\textbf {\bibinfo {volume}
			{13}},\ \bibinfo {pages} {198} (\bibinfo {year}
		{1976}{\natexlab{a}})}\BibitemShut {NoStop}%
	\bibitem [{\citenamefont {Page}(1976{\natexlab{b}})}]{Page1976b}%
	\BibitemOpen
	\bibfield  {author} {\bibinfo {author} {\bibfnamefont {D.~N.}\ \bibnamefont
			{Page}},\ }\href {https://doi.org/10.1103/PhysRevD.14.3260} {\bibfield
		{journal} {\bibinfo  {journal} {Phys. Rev. D}\ }\textbf {\bibinfo {volume}
			{14}},\ \bibinfo {pages} {3260} (\bibinfo {year}
		{1976}{\natexlab{b}})}\BibitemShut {NoStop}%
	\bibitem [{\citenamefont {Gray}\ \emph {et~al.}(2016)\citenamefont {Gray},
		\citenamefont {Schuster}, \citenamefont {Van-Brunt},\ and\ \citenamefont
		{Visser}}]{GrayEtAl2015}%
	\BibitemOpen
	\bibfield  {author} {\bibinfo {author} {\bibfnamefont {F.}~\bibnamefont
			{Gray}}, \bibinfo {author} {\bibfnamefont {S.}~\bibnamefont {Schuster}},
		\bibinfo {author} {\bibfnamefont {A.}~\bibnamefont {Van-Brunt}},\ and\
		\bibinfo {author} {\bibfnamefont {M.}~\bibnamefont {Visser}},\ }\href
	{https://doi.org/10.1088/0264-9381/33/11/115003} {\bibfield  {journal}
		{\bibinfo  {journal} {Class. Quantum Grav.}\ }\textbf {\bibinfo {volume}
			{33}},\ \bibinfo {pages} {115003} (\bibinfo {year} {2016})}\BibitemShut
	{NoStop}%
	\bibitem [{\citenamefont {Chowdhury}\ and\ \citenamefont
		{Banerjee}(2020)}]{ChowdhuryBanerjee2020}%
	\BibitemOpen
	\bibfield  {author} {\bibinfo {author} {\bibfnamefont {A.}~\bibnamefont
			{Chowdhury}}\ and\ \bibinfo {author} {\bibfnamefont {N.}~\bibnamefont
			{Banerjee}},\ }\href {https://doi.org/10.1016/j.physletb.2020.135417}
	{\bibfield  {journal} {\bibinfo  {journal} {Phys. Lett. B}\ }\textbf
		{\bibinfo {volume} {805}},\ \bibinfo {pages} {135417} (\bibinfo {year}
		{2020})}\BibitemShut {NoStop}%
	\bibitem [{\citenamefont {Wei}\ and\ \citenamefont {Liu}(2011)}]{WeiLiu2011}%
	\BibitemOpen
	\bibfield  {author} {\bibinfo {author} {\bibfnamefont {S.-W.}\ \bibnamefont
			{Wei}}\ and\ \bibinfo {author} {\bibfnamefont {Y.-X.}\ \bibnamefont {Liu}},\
	}\href {https://doi.org/10.1103/PhysRevD.84.041501} {\bibfield  {journal}
		{\bibinfo  {journal} {Phys. Rev. D}\ }\textbf {\bibinfo {volume} {84}},\
		\bibinfo {pages} {041501(R)} (\bibinfo {year} {2011})}\BibitemShut {NoStop}%
	\bibitem [{\citenamefont {Wei}\ and\ \citenamefont {Liu}(2013)}]{WeiLiu2013}%
	\BibitemOpen
	\bibfield  {author} {\bibinfo {author} {\bibfnamefont {S.-W.}\ \bibnamefont
			{Wei}}\ and\ \bibinfo {author} {\bibfnamefont {Y.-X.}\ \bibnamefont {Liu}},\
	}\href {https://doi.org/10.1088/1475-7516/2013/11/063} {\bibfield  {journal}
		{\bibinfo  {journal} {JCAP}\ }\textbf {\bibinfo {volume} {2013}}\bibinfo
		{number} { (11)},\ \bibinfo {pages} {063}}\BibitemShut {NoStop}%
	\bibitem [{\citenamefont {D{\'e}canini}\ \emph {et~al.}(2011)\citenamefont
		{D{\'e}canini}, \citenamefont {Esposito-Far{\`e}se},\ and\ \citenamefont
		{Folacci}}]{DecaniniEtAl2011}%
	\BibitemOpen
	\bibfield  {number} {  }\bibfield  {author} {\bibinfo {author} {\bibfnamefont
			{Y.}~\bibnamefont {D{\'e}canini}}, \bibinfo {author} {\bibfnamefont
			{G.}~\bibnamefont {Esposito-Far{\`e}se}},\ and\ \bibinfo {author}
		{\bibfnamefont {A.}~\bibnamefont {Folacci}},\ }\href
	{https://doi.org/10.1103/PhysRevD.83.044032} {\bibfield  {journal} {\bibinfo
			{journal} {Phys. Rev. D}\ }\textbf {\bibinfo {volume} {83}},\ \bibinfo
		{pages} {044032} (\bibinfo {year} {2011})}\BibitemShut {NoStop}%
	\bibitem [{\citenamefont {Konoplya}\ and\ \citenamefont
		{Zhidenko}(2024)}]{KonoplyaZhidenko2024}%
	\BibitemOpen
	\bibfield  {author} {\bibinfo {author} {\bibfnamefont {R.~A.}\ \bibnamefont
			{Konoplya}}\ and\ \bibinfo {author} {\bibfnamefont {A.}~\bibnamefont
			{Zhidenko}},\ }\href {https://doi.org/10.1088/1475-7516/2024/09/068}
	{\bibfield  {journal} {\bibinfo  {journal} {Journal of Cosmology and
				Astroparticle Physics}\ }\textbf {\bibinfo {volume} {2024}}\bibinfo  {number}
		{ (09)},\ \bibinfo {pages} {068}}\BibitemShut {NoStop}%
	\bibitem [{\citenamefont {Konoplya}\ and\ \citenamefont
		{Zhidenko}(2025)}]{KonoplyaZhidenko2025}%
	\BibitemOpen
	\bibfield  {number} {  }\bibfield  {author} {\bibinfo {author} {\bibfnamefont
			{R.~A.}\ \bibnamefont {Konoplya}}\ and\ \bibinfo {author} {\bibfnamefont
			{A.}~\bibnamefont {Zhidenko}},\ }\href
	{https://doi.org/10.1016/j.physletb.2024.139288} {\bibfield  {journal}
		{\bibinfo  {journal} {Phys. Lett. B}\ }\textbf {\bibinfo {volume} {861}},\
		\bibinfo {pages} {139288} (\bibinfo {year} {2025})}\BibitemShut {NoStop}%
	\bibitem [{\citenamefont {Lambiase}\ \emph {et~al.}(2025)\citenamefont
		{Lambiase}, \citenamefont {Gogoi}, \citenamefont {Pantig},\ and\
		\citenamefont {{\"O}vg{\"u}n}}]{LambiaseEtAl2025}%
	\BibitemOpen
	\bibfield  {author} {\bibinfo {author} {\bibfnamefont {G.}~\bibnamefont
			{Lambiase}}, \bibinfo {author} {\bibfnamefont {D.~J.}\ \bibnamefont {Gogoi}},
		\bibinfo {author} {\bibfnamefont {R.~C.}\ \bibnamefont {Pantig}},\ and\
		\bibinfo {author} {\bibfnamefont {A.}~\bibnamefont {{\"O}vg{\"u}n}},\ }\href
	{https://doi.org/10.1016/j.dark.2025.101886} {\bibfield  {journal} {\bibinfo
			{journal} {Phys. Dark Univ.}\ }\textbf {\bibinfo {volume} {49}},\ \bibinfo
		{pages} {101886} (\bibinfo {year} {2025})}\BibitemShut {NoStop}%
	\bibitem [{\citenamefont {Tan}\ \emph {et~al.}(2025)\citenamefont {Tan},
		\citenamefont {Liu}, \citenamefont {Liang},\ and\ \citenamefont
		{Long}}]{TanEtAl2025}%
	\BibitemOpen
	\bibfield  {author} {\bibinfo {author} {\bibfnamefont {Q.}~\bibnamefont
			{Tan}}, \bibinfo {author} {\bibfnamefont {D.}~\bibnamefont {Liu}}, \bibinfo
		{author} {\bibfnamefont {J.}~\bibnamefont {Liang}},\ and\ \bibinfo {author}
		{\bibfnamefont {Z.-W.}\ \bibnamefont {Long}},\ }\href
	{https://doi.org/10.1140/epjc/s10052-025-14407-3} {\bibfield  {journal}
		{\bibinfo  {journal} {Eur. Phys. J. C}\ }\textbf {\bibinfo {volume} {85}},\
		\bibinfo {pages} {687} (\bibinfo {year} {2025})}\BibitemShut {NoStop}%
	\bibitem [{\citenamefont {Ahmed}\ \emph
		{et~al.}(2026{\natexlab{d}})\citenamefont {Ahmed}, \citenamefont {Fathi},\
		and\ \citenamefont {Silva}}]{Ahmed2026ChargedRegularBH}%
	\BibitemOpen
	\bibfield  {author} {\bibinfo {author} {\bibfnamefont {F.}~\bibnamefont
			{Ahmed}}, \bibinfo {author} {\bibfnamefont {M.}~\bibnamefont {Fathi}},\ and\
		\bibinfo {author} {\bibfnamefont {E.~O.}\ \bibnamefont {Silva}},\ }\href
	{https://arxiv.org/abs/2604.11357} {} (\bibinfo {year}
	{2026}{\natexlab{d}}),\ \Eprint {https://arxiv.org/abs/2604.11357}
	{arXiv:2604.11357 [gr-qc]} \BibitemShut {NoStop}%
	\bibitem [{\citenamefont {Heisenberg}\ and\ \citenamefont
		{Euler}(1936{\natexlab{b}})}]{Heisenberg1936}%
	\BibitemOpen
	\bibfield  {author} {\bibinfo {author} {\bibfnamefont {W.}~\bibnamefont
			{Heisenberg}}\ and\ \bibinfo {author} {\bibfnamefont {H.}~\bibnamefont
			{Euler}},\ }\href {https://doi.org/10.1007/BF01343663} {\bibfield  {journal}
		{\bibinfo  {journal} {Zeitsch. für Phys.}\ }\textbf {\bibinfo {volume}
			{98}},\ \bibinfo {pages} {714} (\bibinfo {year}
		{1936}{\natexlab{b}})}\BibitemShut {NoStop}%
	\bibitem [{\citenamefont {Yajima}\ and\ \citenamefont
		{Tamaki}(2001{\natexlab{b}})}]{Yajima2001}%
	\BibitemOpen
	\bibfield  {author} {\bibinfo {author} {\bibfnamefont {H.}~\bibnamefont
			{Yajima}}\ and\ \bibinfo {author} {\bibfnamefont {T.}~\bibnamefont
			{Tamaki}},\ }\href {https://doi.org/10.1103/PhysRevD.63.064007} {\bibfield
		{journal} {\bibinfo  {journal} {Phys. Rev. D}\ }\textbf {\bibinfo {volume}
			{63}},\ \bibinfo {pages} {064007} (\bibinfo {year}
		{2001}{\natexlab{b}})}\BibitemShut {NoStop}%
	\bibitem [{\citenamefont {Salazar}\ \emph
		{et~al.}(1987{\natexlab{b}})\citenamefont {Salazar}, \citenamefont
		{Garcia~D.},\ and\ \citenamefont {Plebanski}}]{Salazar1987}%
	\BibitemOpen
	\bibfield  {author} {\bibinfo {author} {\bibfnamefont {H.}~\bibnamefont
			{Salazar}}, \bibinfo {author} {\bibfnamefont {A.}~\bibnamefont {Garcia~D.}},\
		and\ \bibinfo {author} {\bibfnamefont {J.}~\bibnamefont {Plebanski}},\ }\href
	{https://doi.org/10.1063/1.527570} {\bibfield  {journal} {\bibinfo  {journal}
			{J. Math. Phys.}\ }\textbf {\bibinfo {volume} {28}},\ \bibinfo {pages} {2171}
		(\bibinfo {year} {1987}{\natexlab{b}})}\BibitemShut {NoStop}%
	\bibitem [{\citenamefont {Magos}\ and\ \citenamefont
		{Breton}(2020)}]{Magos2020}%
	\BibitemOpen
	\bibfield  {author} {\bibinfo {author} {\bibfnamefont {D.}~\bibnamefont
			{Magos}}\ and\ \bibinfo {author} {\bibfnamefont {N.}~\bibnamefont {Breton}},\
	}\href {https://doi.org/10.1103/PhysRevD.102.084011} {\bibfield  {journal}
		{\bibinfo  {journal} {Phys. Rev. D}\ }\textbf {\bibinfo {volume} {102}},\
		\bibinfo {pages} {084011} (\bibinfo {year} {2020})}\BibitemShut {NoStop}%
	\bibitem [{\citenamefont {Li}\ and\ \citenamefont
		{Yang}(2012{\natexlab{b}})}]{Li2012}%
	\BibitemOpen
	\bibfield  {author} {\bibinfo {author} {\bibfnamefont {M.~H.}\ \bibnamefont
			{Li}}\ and\ \bibinfo {author} {\bibfnamefont {K.~C.}\ \bibnamefont {Yang}},\
	}\href {https://doi.org/10.1103/PhysRevD.86.123015} {\bibfield  {journal}
		{\bibinfo  {journal} {Phys. Rev. D}\ }\textbf {\bibinfo {volume} {86}},\
		\bibinfo {pages} {123015} (\bibinfo {year} {2012}{\natexlab{b}})}\BibitemShut
	{NoStop}%
	\bibitem [{\citenamefont {Xu}\ \emph {et~al.}(2017)\citenamefont {Xu},
		\citenamefont {Hou}, \citenamefont {Wang},\ and\ \citenamefont
		{Liao}}]{Xu2017}%
	\BibitemOpen
	\bibfield  {author} {\bibinfo {author} {\bibfnamefont {Z.}~\bibnamefont
			{Xu}}, \bibinfo {author} {\bibfnamefont {X.}~\bibnamefont {Hou}}, \bibinfo
		{author} {\bibfnamefont {J.}~\bibnamefont {Wang}},\ and\ \bibinfo {author}
		{\bibfnamefont {Y.}~\bibnamefont {Liao}},\ }\href
	{https://doi.org/10.1155/2017/8079280} {\bibfield  {journal} {\bibinfo
			{journal} {Adv. High Energy Phys.}\ }\textbf {\bibinfo {volume} {2017}},\
		\bibinfo {pages} {1} (\bibinfo {year} {2017})}\BibitemShut {NoStop}%
	\bibitem [{\citenamefont {Konoplya}\ and\ \citenamefont
		{Zhidenko}(2011)}]{KonoplyaZhidenko2011}%
	\BibitemOpen
	\bibfield  {author} {\bibinfo {author} {\bibfnamefont {R.~A.}\ \bibnamefont
			{Konoplya}}\ and\ \bibinfo {author} {\bibfnamefont {A.}~\bibnamefont
			{Zhidenko}},\ }\href {https://doi.org/10.1103/RevModPhys.83.793} {\bibfield
		{journal} {\bibinfo  {journal} {Rev. Mod. Phys.}\ }\textbf {\bibinfo {volume}
			{83}},\ \bibinfo {pages} {793} (\bibinfo {year} {2011})}\BibitemShut
	{NoStop}%
	\bibitem [{\citenamefont {Konoplya}\ \emph {et~al.}(2019)\citenamefont
		{Konoplya}, \citenamefont {Zhidenko},\ and\ \citenamefont
		{Zinhailo}}]{KonoplyaZhidenkoZinhailo2019}%
	\BibitemOpen
	\bibfield  {author} {\bibinfo {author} {\bibfnamefont {R.~A.}\ \bibnamefont
			{Konoplya}}, \bibinfo {author} {\bibfnamefont {A.}~\bibnamefont {Zhidenko}},\
		and\ \bibinfo {author} {\bibfnamefont {A.~F.}\ \bibnamefont {Zinhailo}},\
	}\href {https://doi.org/10.1088/1361-6382/ab2735} {\bibfield  {journal}
		{\bibinfo  {journal} {Class. Quantum Grav.}\ }\textbf {\bibinfo {volume}
			{36}},\ \bibinfo {pages} {155002} (\bibinfo {year} {2019})}\BibitemShut
	{NoStop}%
	\bibitem [{\citenamefont {Mashhoon}(1985)}]{Mashhoon1985}%
	\BibitemOpen
	\bibfield  {author} {\bibinfo {author} {\bibfnamefont {B.}~\bibnamefont
			{Mashhoon}},\ }\href {https://doi.org/10.1103/PhysRevD.31.290} {\bibfield
		{journal} {\bibinfo  {journal} {Phys. Rev. D}\ }\textbf {\bibinfo {volume}
			{31}},\ \bibinfo {pages} {290} (\bibinfo {year} {1985})}\BibitemShut
	{NoStop}%
	\bibitem [{\citenamefont {Hawking}(1974)}]{Hawking1974}%
	\BibitemOpen
	\bibfield  {author} {\bibinfo {author} {\bibfnamefont {S.~W.}\ \bibnamefont
			{Hawking}},\ }\href {https://doi.org/10.1038/248030a0} {\bibfield  {journal}
		{\bibinfo  {journal} {Nature}\ }\textbf {\bibinfo {volume} {248}},\ \bibinfo
		{pages} {30} (\bibinfo {year} {1974})}\BibitemShut {NoStop}%
	\bibitem [{\citenamefont {Mashhoon}(1973)}]{Mashhoon1973}%
	\BibitemOpen
	\bibfield  {author} {\bibinfo {author} {\bibfnamefont {B.}~\bibnamefont
			{Mashhoon}},\ }\href {https://doi.org/10.1103/PhysRevD.7.2807} {\bibfield
		{journal} {\bibinfo  {journal} {Phys. Rev. D}\ }\textbf {\bibinfo {volume}
			{7}},\ \bibinfo {pages} {2807} (\bibinfo {year} {1973})}\BibitemShut
	{NoStop}%
	\bibitem [{\citenamefont {Misner}\ \emph {et~al.}(1973)\citenamefont {Misner},
		\citenamefont {Thorne},\ and\ \citenamefont {Wheeler}}]{Misner1973}%
	\BibitemOpen
	\bibfield  {author} {\bibinfo {author} {\bibfnamefont {C.~W.}\ \bibnamefont
			{Misner}}, \bibinfo {author} {\bibfnamefont {K.~S.}\ \bibnamefont {Thorne}},\
		and\ \bibinfo {author} {\bibfnamefont {J.~A.}\ \bibnamefont {Wheeler}},\
	}\href@noop {} {\emph {\bibinfo {title} {Gravitation}}}\ (\bibinfo
	{publisher} {W. H. Freeman},\ \bibinfo {address} {San Francisco},\ \bibinfo
	{year} {1973})\BibitemShut {NoStop}%
\end{thebibliography}
\end{document}